\documentclass[pra,showpacs,showkeywords,graphics,epsfigure,superscriptaddress]{revtex4}

\usepackage{color}
\usepackage{esvect}
\usepackage{graphicx}
\usepackage{graphics}
\usepackage{bbm}
\usepackage{amsmath}
\usepackage{amsfonts}
\usepackage{amssymb}
\usepackage{latexsym}

\usepackage{braket}
\usepackage{amsbsy}
\usepackage{amsthm}
\usepackage{bm}
\usepackage{dsfont} 

\newcommand{\beq}{\begin{eqnarray}}
\newcommand{\eeq}{\end{eqnarray}}

\newtheorem{theorem}{Theorem}

\newcommand{\A}{\mathcal{A}}
\newcommand{\B}{\mathcal{B}}

\begin{document}

\title{Entangled Entanglement: The Geometry of GHZ States}
\author{Gabriele Uchida}
\address{University of Vienna, Faculty of Computer Science, W\" ahringer Strasse 29, 1090 Vienna, Austria}
\author{Reinhold A. Bertlmann}
\address{University of Vienna, Faculty of Physics, Boltzmanngasse 5, 1090 Vienna, Austria}
\author{Beatrix C. Hiesmayr}
\address{University of Vienna, Faculty of Physics, Boltzmanngasse 5, 1090 Vienna, Austria}
\affiliation{
Beatrix.Hiesmayr@univie.ac.at}

\begin{abstract}
The familiar Greenberger-Horne-Zeilinger (GHZ) states can be rewritten by entangling the Bell states for two qubits with a state of the third qubit, which is dubbed \emph{entangled entanglement}. We show that in this way we obtain all $8$ independent GHZ states that form the simplex of entangled entanglement, the \emph{magic simplex}. The construction procedure allows a generalization to higher dimensions both, in the degrees of freedom (considering qudits) as well as in the number of particles (considering $n$-partite states). Such bases of GHZ-type states exhibit a certain geometry that is relevant for experimental and quantum information theoretic applications. Furthermore, we study the geometry of these particular state spaces, the inherent symmetries, the cyclicity of the phase operations, and the regions of (genuine multi-partite) entanglement and the several classes of separability. We find non-trivial geometrical properties and a conceptually clear procedure to compare state spaces of different dimensions and number of particles.
\end{abstract}



\maketitle


\section{Introduction}

Entanglement is one of the most fundamental and fascinating nonclassical phenomena in quantum physics, see e.g. Ref.~\cite{bertlmann-zeilinger02}. In low dimensions, where the Hilbert space is a tensor product of dimension $2 \otimes 2\,$, two qubits are either entangled or separable and the geometric structure of the space is well understood. However, in higher dimensions the geometry becomes much richer~\cite{horodecki99, baumgartner-hiesmayr-narnhofer06, baumgartner-hiesmayr-narnhofer07, baumgartner-hiesmayr-narnhofer08, bertlmann-krammer-AnnPhys09, bertlmann-krammer-PRA-78-08, bertlmann-krammer-PRA-77-08,Chruscinski1,Chruscinski2}, new features show up, e.g., bound entanglement, entanglement that cannot be distilled. Only recently bipartite bound entanglement has been experimentally revealed by exploring photons entangled in their orbital angular momentum degrees of freedom~\cite{hiesmayrloeffler1,hiesmayrloeffler2}. A complete and systematic investigation for such generalized higher dimensional cases is still missing.

Another generalization is obtained by increasing the number of particles $n$. Here further peculiarities occur like bi-separability, $k$-separability in general, or genuine multi-partite entanglement, see e.g. the reviews \cite{HHHH07,gmedet3}. The reason is that the algebra of the density matrix that corresponds to a quantum state can be factorized in different ways \cite{TBKN,zanardi2001}. Such factorizations, however, are not unique.  Moreover, if no bi-partition is possible the state can be still entangled in many different physical ways and so far only necessary but not sufficient criteria have been developed, e.g. Refs.~\cite{hmgh,GHZpaper1,GHZpaper2,Mario,Szalay,Chruscinski1,Chruscinski2}.

An amazing feature that occurs in $n$-partite systems is \emph{entangled entanglement}. The term was coined by Krenn and Zeilinger \cite{Krenn-Zeilinger} to characterize the phenomenon that the entanglement of two qubits, expressed by the Bell states, can be entangled further with a third qubit, producing such a particular Greenberger-Horne-Zeilinger (GHZ) state. We take up this idea, develop it further and show that all eight independent (maximally entangled) GHZ states, can be expressed, geometrically quite obviously, in an entangled entanglement form. These eight states configure the so-called \emph{magic simplex}~\cite{baumgartner-hiesmayr-narnhofer06}. Our construction procedure, which is entirely systematic, can be easily generalized to higher dimensions $d$ and to any finite number of particles $n$, namely to $n$-partite qudit states $d\otimes d\otimes d\otimes d \otimes ...\otimes d=d^{\otimes n}$.

To obtain an understanding for the entangled entanglement phenomenon we discuss the case of GHZ in Sect.~\ref{sec:physics and mathematics}, the physical aspect with respect to the Einstein-Podolsky-Rosen (EPR) paradox and the mathematical structure, the freedom to factorize a tensor product of algebras (or Hilbert spaces) in different ways, which forms the mathematical basis for the phenomenon of entangled entanglement. In Sect.~\ref{sec:entangled entanglement construction} we introduce our procedure how to construct systematically the states of entangled entanglement for any higher dimension. The use of Weyl operators turns out to be very helpful. We also show the geometric structure of the space, the symmetries inherent in a magic simplex and the cyclicity of the phase operations, when moving from one simplex to another. Sect.~\ref{trafoSepEntangle} presents our interferometric point of view of the transformations between separability and entanglement. The nature of the quantum states inside the simplex, the regions of the several kinds of separability and entanglement is investigated in Sect.~\ref{sec:separability multipartite entanglement}. The entanglement criterion established in Ref.~\cite{hmgh} turns out to be very useful to determine the different regions of separability and genuine multi-partite entanglement of the quantum states. We also compare the geometry in different dimensions and number of particles, an advantage of our construction. Finally, conclusions are drawn in Sect.~\ref{sec:conclusions} emphasising how the developed toolbox will help in future to explore certain quantum phenomena.


\section{Physical Aspect and Mathematical Structure}\label{sec:physics and mathematics}

Let us consider the well-known GHZ state~\cite{GHZ-paper, GHZ-paper2}
\begin{equation}\label{GHZ1minus RRR+LLL}
\ket{GHZ1^-}_{123} \;=\; \frac{1}{\sqrt{2}}\,\big(\ket{R}_1\otimes\ket{R}_2\otimes\ket{R}_3\,+\,\ket{L}_1\otimes\ket{L}_2\otimes\ket{L}_3\big)\\[3pt]\;,
\end{equation}
where $\ket{R}, \ket{L}$ denote the right and left handed circularly polarized photons. Interestingly, expression (\ref{GHZ1minus RRR+LLL}) can be re-expressed by decomposing (\ref{GHZ1minus RRR+LLL}) into linearly polarized states $\ket{H}, \ket{V}$ and Bell states
\begin{equation}\label{GHZ1minus HVBell}
\ket{GHZ1^-}_{123} \;=\; \frac{1}{\sqrt{2}}\,\big(\ket{H}_1\otimes\ket{\phi^-}_{23}\,-\,\ket{V}_1\otimes\ket{\psi^+}_{23}\big)\;,
\end{equation}
where $\ket{\phi^{\pm}} \,=\, \frac{1}{\sqrt{2}}\,\big(\ket{H}\otimes\ket{H} \,\pm\, \ket{V}\otimes\ket{V}\big) \,,\; \ket{\psi^{\pm}} \,=\, \frac{1}{\sqrt{2}}\,\big(\ket{H}\otimes\ket{V} \,\pm\, \ket{V}\otimes\ket{H}\big)$ represent the familiar maximally entangled Bell states. Recall that the linearly polarized states $|H/V\rangle$ are related to the circularly polarized states via $|R/L\rangle=\frac{1}{\sqrt{2}}(|H\rangle\pm i |V\rangle)$.

The GHZ state as expressed in Eq.~(\ref{GHZ1minus HVBell}) obviously represents entangled entanglement. This feature has been verified experimentally by Zeilinger's group~\cite{Walther-Resch-Brukner-Zeilinger} who has performed a Bell-type experiment on three particles, where one part, Alice on line 1, projects onto the horizontally $\ket{H}_{1}$ or vertically $\ket{V}_{1}$ polarized state and the other part, Bob on lines 2 and 3, projects onto the maximally entangled states $\ket{\phi^-}_{23}$ or $\ket{\psi^+}_{23}$. Then the authors test a Clauser-Horne-Shimony-Holt inequality established between Alice and Bob and find a strong violation of the inequality (of the Bell parameter) by more than 5 standard deviations. Thus the entangled Bell states of the two photons of Bob are definitely entangled again with the single photon of Alice.

\begin{figure}
	\centering
	(a)\includegraphics[width=0.4\textwidth]{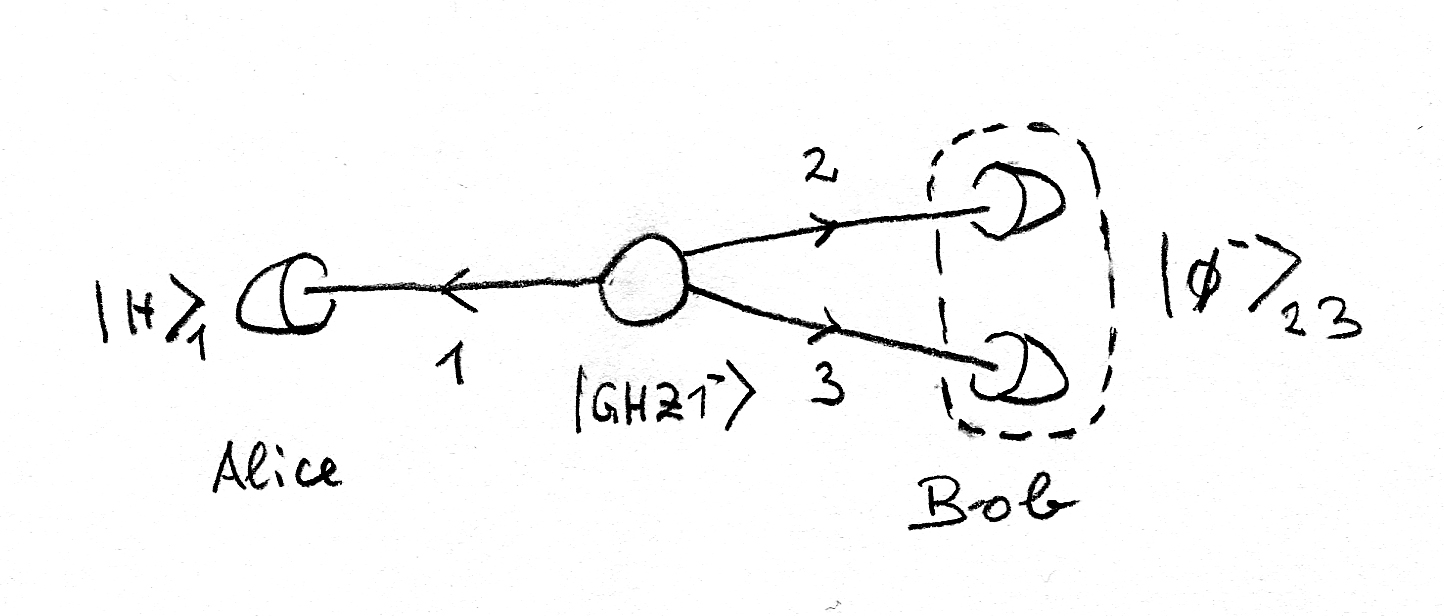}
(b)\includegraphics[width=0.4\textwidth]{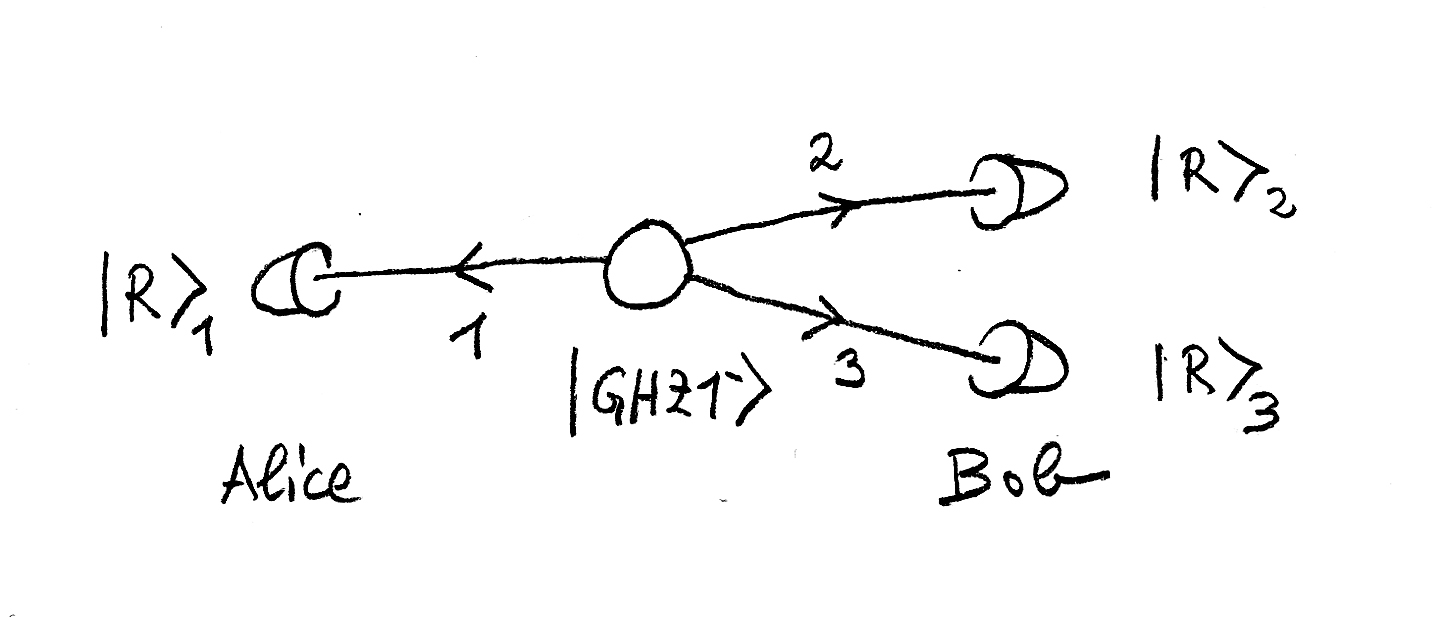}
	\caption{(a) Bob's photons are in an entangled state. (b) Bob's photons are in a separable state.}
	\label{fig:Bob photons entangled}
\end{figure}

What is the physical significance of it, in particular, in the light of an EPR reasoning? Let us start with an EPR-like discussion as in Ref.~\cite{Walther-Resch-Brukner-Zeilinger}. If Alice is measuring the linearly polarized state $\ket{H}_{1}$ then Bob will find the Bell state $\ket{\phi^-}_{23}$ for his two photons (see Fig.~\ref{fig:Bob photons entangled}~(a)). If she obtains a $\ket{V}_{1}$ state in her measurement then Bob will get the Bell state $\ket{\psi^+}_{23}\,$. This perfect correlation between the polarization state of one photon on Alice's side and the entangled state of the two photons on Bob's side implies, under the EPR premises of realism and no action at a distance, that the entangled state of the two photons must represent an element of reality. Whereas the individual photons of Bob, which have no well-defined property, do not correspond to such elements. For a realist this is a surprising feature, indeed.

If, on the other hand, Alice is measuring a right-handed circularly polarized state $\ket{R}_1$ then Bob will find his two photons in a separable state $\ket{R}_2 \otimes \ket{R}_3$ or if Alice measures $\ket{L}_1$ Bob will get $\ket{L}_2 \otimes \ket{L}_3\,$ (see Fig.~\ref{fig:Bob photons entangled}~(b)). Then the two photons of Bob contain individually an element of reality, which is more satisfactory for a realist. Thus by the specific kind of measurement, projecting on linearly or circularly polarized photons, Alice is able to switch on Bob's side the properties of the two photons---and their reality content---between entanglement and separability.\\

Even more puzzling becomes this feature when one thinks of entanglement of internal and external degrees of freedom which can also be experimentally achieved, e.g. with single neutrons propagating through an interferometer. In Ref.~\cite{neutron} experimenters proved that a GHZ-like state for single neutrons entangled in path-spin-energy can be produced where the same above considerations have to hold!\\
\\
\textit{How can we understand this switching phenomenon between entanglement and separability?}\\
\\
A quantum theorist can trace this switch back to two different factorizations of the tensor product of three algebras $\A_1 \otimes \A_2 \otimes \A_3 \,$, where $\A_1$ belongs to Alice and $\A_2 \otimes \A_3$ to Bob. There is total democracy between the different factorizations~\cite{TBKN,zanardi2001}, no partition has ontologically a superior status over any other one. For an experimentalist, however, a certain factorization is preferred and is clearly fixed by the set-up.

For bipartite states there exists the following theorem~\cite{TBKN}, where $\rho\,=\,\left|\,\psi\,\right\rangle\left\langle\,\psi\,\right|$ denotes the corresponding density matrix of the quantum state $\ket{\psi}$:

\begin{theorem}[Factorization algebra]\ \
For any pure state $\rho$ one can find a factorization $M \,=\, \A_1 \otimes \A_2$ such that $\rho$ is separable with respect to this factorization and another factorization $M \,=\, \B_1 \otimes \B_2$ where $\rho$ appears to be maximally entangled.
\label{theorem:factorization-algebra pure states}
\end{theorem}

\noindent\textbf{Example I:}\ \ \
To illustrate Theorem~\ref{theorem:factorization-algebra pure states} we consider the circularly polarized states $\{\ket{R}, \ket{L}\}$ since we will need their transformations in Example $I\!I$ and in Sect.~\ref{trafoSepEntangle}. Generally, we find the following structure for the unitary matrix $U$ that transform separable states $\{\ket{R}\otimes\ket{R} , \ket{L}\otimes\ket{L}\}$ into maximally entangled Bell states $\{\ket{\phi^-} , \ket{\psi^+}\}$
\begin{equation}\label{unitary V for Bell states}
U \ket{R} \otimes \ket{R} \;=\; - \ket{\phi^-} \qquad \mbox{and} \qquad U \ket{L} \otimes \ket{L} \;=\; i \ket{\psi^+} \,,
\end{equation}
where the unitary matrix $U$ has the structure
\beq
U=(U_{BS}\otimes U_{BS})\cdot (U_{phase}\otimes &&U_{phase})^\dagger\cdot U_{ent}\cdot (U_{phase}\otimes U_{phase})\cdot (U_{BS}\otimes U_{BS})\;.
\eeq
The unitary matrix $U_{ent}$ transforms $\ket{H}\otimes\ket{H}$ or $\ket{V}\otimes\ket{V}$ into the Bell states $|\phi^-\rangle$ and $|\psi^+\rangle$, respectively, whereas the mixed tensor products $\ket{H}\otimes\ket{V}$ and $\ket{V}\otimes\ket{H}$ are eigenstates of $U_{ent}$. The index $BS$ stands for beam splitter and $U_{phase}$ is a phase transformation of $\frac{\pi}{2}$ in both interfering ``beams''. All unitary operators are explained and given explicitly in Sect.~\ref{trafoSepEntangle}.\\


In case of GHZ states, which are defined on a tensor product of three algebras, an analogous theorem holds:

\begin{theorem}[Factorization algebra]\ \
For any pure GHZ state $\rho$ one can find a factorization $M \,=\, \A_1 \otimes \A_2 \otimes \A_3$ such that $\rho$ is separable with respect to this factorization and another factorization $M \,=\, \B_1 \otimes \B_2 \otimes \B_3$ where $\rho$ appears to be maximally entangled.
\label{theorem:factorization-algebra GHZ pure states}
\end{theorem}

\noindent\textbf{Example $I\!I$:}\ \ \
For the extended structure, the tensor product of three algebras, the following unitary matrix $U$ exemplifies Theorem~\ref{theorem:factorization-algebra GHZ pure states}. The matrix $U$ transforms, for example, the separable state $\ket{R}_1\otimes\ket{R}_2\otimes\ket{R}_3\,$ into the entangled state $\ket{GHZ1^-}_{123}$
\begin{equation}\label{unitary matrix GHZ to RRR}
U \ket{R}_1\otimes\ket{R}_2\otimes\ket{R}_3 \;=\; \ket{GHZ1^-}_{123}
\end{equation}
with
\beq
U=U_{BS}^{\otimes 3}\cdot (U_{phase}^{\otimes 3})^\dagger \cdot U_{ent}\cdot U_{phase}^{\otimes 3}\cdot U_{BS}^{\otimes 3}\;.
\eeq

Having found the structure of entangled entanglement, it is quite natural to ask if other GHZ states can be expressed in a similar way. Yes, it can be done, we can construct a complete orthonormal system (CONS). Recalling the geometry of qubits in $2 \otimes 2$, the tetrahedron of the Bell states \cite{bertlmann-narnhofer-thirring02, vollbrecht-werner-PRA00, horodecki-R-M96} and its possible extensions to qudits and more particles \cite{baumgartner-hiesmayr-narnhofer06, baumgartner-hiesmayr-narnhofer07, baumgartner-hiesmayr-narnhofer08, bertlmann-krammer-AnnPhys09, bertlmann-krammer-PRA-77-08, bertlmann-krammer-PRA-78-08}, it is quite obvious how to proceed. We just have to entangle the geometrically opposite states $\ket{\phi^-}$ and $\ket{\psi^+}$ or $\ket{\phi^+}$ and $\ket{\psi^-}$ together with $\ket{H}$ and $\ket{V}$ and respect the symmetric and antisymmetric property respectively. In this way we immediately find an orthonormal basis of eight states
\begin{equation}\label{GHZ basis with Bell states}
\begin{array}{rcl}
\ket{GHZ1^+}_{123}&=&\frac{1}{\sqrt{2}}\,\big(\ket{H}_1\otimes\ket{\phi^-}_{23}\,+\,\ket{V}_1\otimes\ket{\psi^+}_{23}\big)\\
\ket{GHZ1^-}_{123}&=&\frac{1}{\sqrt{2}}\,\big(\ket{H}_1\otimes\ket{\phi^-}_{23}\,-\,\ket{V}_1\otimes\ket{\psi^+}_{23}\big)\\
\ket{GHZ2^+}_{123}&=&\frac{1}{\sqrt{2}}\,\big(\ket{H}_1\otimes\ket{\phi^+}_{23}\,+\,\ket{V}_1\otimes\ket{\psi^-}_{23}\big)\\
\ket{GHZ2^-}_{123}&=&\frac{1}{\sqrt{2}}\,\big(\ket{H}_1\otimes\ket{\phi^+}_{23}\,-\,\ket{V}_1\otimes\ket{\psi^-}_{23}\big)\\
\ket{GHZ3^+}_{123}&=&\frac{1}{\sqrt{2}}\,\big(\ket{V}_1\otimes\ket{\phi^-}_{23}\,+\,\ket{H}_1\otimes\ket{\psi^+}_{23}\big)\\
\ket{GHZ3^-}_{123}&=&\frac{1}{\sqrt{2}}\,\big(\ket{V}_1\otimes\ket{\phi^-}_{23}\,-\,\ket{H}_1\otimes\ket{\psi^+}_{23}\big)\\
\ket{GHZ4^+}_{123}&=&\frac{1}{\sqrt{2}}\,\big(\ket{V}_1\otimes\ket{\phi^+}_{23}\,+\,\ket{H}_1\otimes\ket{\psi^-}_{23}\big)\\
\ket{GHZ4^-}_{123}&=&\frac{1}{\sqrt{2}}\,\big(\ket{V}_1\otimes\ket{\phi^+}_{23}\,-\,\ket{H}_1\otimes\ket{\psi^-}_{23}\big)\;.
\end{array}
\end{equation}

These eight states form the vertices of the magic simplex $\mathbbm{S}$ in the corresponding eight dimensional Hilbert space of the tensor product $2\otimes 2\otimes 2\,$. It is the analogue to the tetrahedron of Bell states in $2\otimes 2\,$ dimensions. The set $\mathbbm{S}$ itself consists of the convex combinations of all the corresponding density matrices $\rho_{GHZi^\pm}$
\beq\label{magic-simplex}
&&{\emph{Magic simplex of entangled entanglement:}}\nonumber\\
&&\mathbbm{S} \;:=\; \{ \,\rho= \sum_{i=1,..4;k=+,-} \lambda_i^k \;\rho_{GHZi^k}| \quad \lambda_i^{\pm} \ge 0,\quad \sum  \lambda_i^{\pm} = 1\,\}\;,
\eeq
where
\begin{equation}
\rho_{GHZi^\pm} \;=\; \Ket{GHZi^\pm}\Bra{GHZi^\pm}, \quad i=1,\ldots,4 \;.
\end{equation}

The convex combination (\ref{magic-simplex}) of the GHZ states builds up a simplex with the maximally mixed state $\frac{1}{8} \mathbbm{1}$ in its center. All the density matrices inside this simplex represent valid quantum states: The eigenvalues of every state
$\rho=\sum_{i=1,..4;k=+,-} \lambda_i^k \,\rho_{GHZi^k}$ in the simplex are just equal to the eight coefficients $\lambda_i^k$ \cite{Uchida-DA}. For the elements of $\mathbbm{S}$, we know, by construction, that $ \lambda_i^{\pm} \ge 0$ and  $\sum  \lambda_i^{\pm} = 1\,$,
so that inside the simplex all the eigenvalues are non-negative and for the trace we have ${\rm{Tr}} \rho = 1\,$. Therefore $\rho \in \mathbbm{S} $
is a valid density matrix.

On the other hand, we can show that the matrices outside the simplex $\mathbbm{S}$ have to violate one of the above properties and
therefore do not form density matrices of physical states. Therefore we have ensured that in the corresponding Hilbert-Schmidt space the valid
density matrices are exactly represented by elements of the simplex $\mathbbm{S}$.\\

\noindent\textbf{Example $I\!I\!I$:}\ \ \
Let us illustrate above statements by considering the boundary of the simplex. For example, we choose density matrices on the following one-dimensional facet
\begin{equation}
\rho_{Facet_{\alpha}1^+1^-} \;=\; \alpha  \rho_{GHZ1^+} \,+\, (1-\alpha) \rho_{GHZ1^-}, \quad 0\le \alpha \le 1 \;.
\end{equation}
Then the density matrix, a state on a ray from the maximally mixed to a state on the facet,
\begin{equation}
\rho \;=\; \frac{1}{8}\,\mathbbm{1} \,+\, \mu \,(\rho_{Facet_{\alpha}1^+1^-} \,-\, \frac{1}{8}\,\mathbbm{1}) \;=\;
\mu \,\rho_{Facet_{\alpha}1^+1^-} \,+\, (1-\mu) \,\frac{1}{8}\,\mathbbm{1} \quad\mbox{with}\quad \mu \ge 0
\end{equation}
has eigenvalues $\frac{1-\mu}{8}$ (6 times), $\frac{1+7\mu-8\alpha \mu}{8}$ and $\frac{1-\mu+8\alpha \mu}{8}$ .

\begin{figure}
	\centering
	\includegraphics[width=0.4\textwidth]{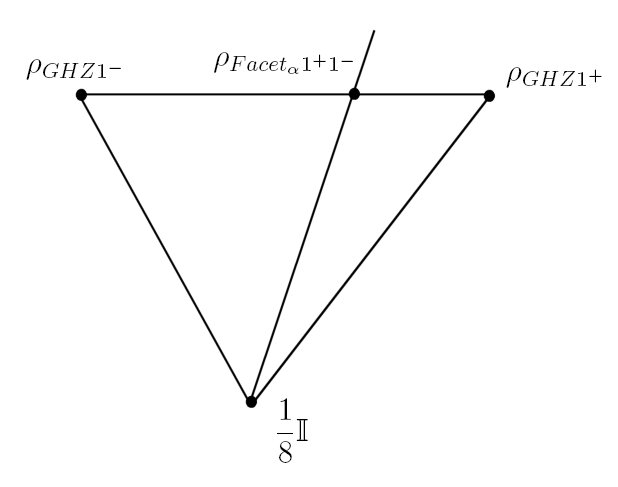}
	\caption{Illustration -- projection into a subspace -- for states in- and outside the magic simplex $\mathbbm{S}$.}
	\label{fig:simplex facet}
\end{figure}

Obviously all the eigenvalues are nonnegative for $\mu \le 1$, but crossing the boundary of $\mathbbm{S}$ by setting $\mu >1\,$, we obtain
negative eigenvalues since then $\frac{1-\mu}{8}$ is obviously negative, that is, the states are not any more physical (see Fig.~\ref{fig:simplex facet}).\\

All states of entangled entanglement (\ref{GHZ basis with Bell states}) can certainly be re-expressed by tensor products of right handed $\ket{R}$ and left handed $\ket{L}$ circularly polarized photon states \cite{Uchida-DA}
\begin{equation}
\begin{array}{rcl}\label{GHZ basis with RL states}
\Ket{GHZ1^+}_{123}&=&\frac{1}{\sqrt{2}}\,\big(\Ket{R}_1\otimes\Ket{L}_2\otimes\Ket{L}_3\,+\,\Ket{L}_1\otimes\Ket{R}_2\otimes\Ket{R}_3\big)\\[3pt]
\Ket{GHZ1^-}_{123}&=&\frac{1}{\sqrt{2}}\,\big(\Ket{R}_1\otimes\Ket{R}_2\otimes\Ket{R}_3\,+\,\Ket{L}_1\otimes\Ket{L}_2\otimes\Ket{L}_3\big)\\[3pt]
\Ket{GHZ2^+}_{123}&=&\frac{1}{\sqrt{2}}\,\big(\Ket{R}_1\otimes\Ket{R}_2\otimes\Ket{L}_3\,+\,\Ket{L}_1\otimes\Ket{L}_2\otimes\Ket{R}_3\big)\\[3pt]
\Ket{GHZ2^-}_{123}&=&\frac{1}{\sqrt{2}}\,\big(\Ket{R}_1\otimes\Ket{L}_2\otimes\Ket{R}_3\,+\,\Ket{L}_1\otimes\Ket{R}_2\otimes\Ket{L}_3\big)\\[3pt]
\Ket{GHZ3^+}_{123}&=&-\frac{i}{\sqrt{2}}\,\big(\Ket{R}_1\otimes\Ket{R}_2\otimes\Ket{R}_3\,-\,\Ket{L}_1\otimes\Ket{L}_2\otimes\Ket{L}_3\big)\\[3pt]
\Ket{GHZ3^-}_{123}&=&-\frac{i}{\sqrt{2}}\,\big(\Ket{R}_1\otimes\Ket{L}_2\otimes\Ket{L}_3\,-\,\Ket{L}_1\otimes\Ket{L}_2\otimes\Ket{R}_3\big)\\[3pt]
\Ket{GHZ4^+}_{123}&=&-\frac{i}{\sqrt{2}}\,\big(\Ket{R}_1\otimes\Ket{L}_2\otimes\Ket{R}_3\,-\,\Ket{L}_1\otimes\Ket{R}_2\otimes\Ket{L}_3\big)\\[3pt]
\Ket{GHZ4^-}_{123}&=&-\frac{i}{\sqrt{2}}\,\big(\Ket{R}_1\otimes\Ket{R}_2\otimes\Ket{L}_3\,-\,\Ket{L}_1\otimes\Ket{L}_2\otimes\Ket{R}_3\big)\;.
\end{array}
\end{equation}
Of course, via local unitary transformations the eight GHZ states (\ref{GHZ basis with RL states}) can be transformed into the corresponding states containing only linearly polarized $\Ket{H}$ and $\Ket{V}$ states.\\

The appeal and importance of the construction of the entangled entanglement is that this procedure can be easily generalized to construct the corresponding states of higher dimensions $d$ and arbitrary number of particles $n$. Entangling the GHZ states (\ref{GHZ basis with Bell states}) again with $\Ket{H}$ and $\Ket{V}$ we obtain the corresponding simplex in the $2\otimes 2\otimes 2\otimes 2$ tensor space and so on. In this way we can construct all higher dimensional simplices of entangled entanglement states in a straightforward way just by entangling again the vertices of the simplex with $\Ket{H}$ and $\Ket{V}$. Thus we find the magic simplex for any particle number. The extension to higher dimensional system is obtained by generalization of the Pauli matrices to the unitary Wely operators. The construction procedure, where the Weyl operators simplify the method, and the investigation of the geometry of these simplices, the regions of entanglement and separability, we are going to present in the following sections.


\section{Entangled Entanglement: Construction of GHZ States}\label{sec:entangled entanglement construction}

Let us first investigate two-dimensional states, i.e. qubits.
Without loss of generality we start with the following seed state
\begin{eqnarray}
|\Phi_1\rangle = |0\rangle.
\end{eqnarray}

In our presentation we will use the convenient notation $|0\rangle$ and $|1\rangle$ of quantum information instead of $|H\rangle$ and $|V\rangle$, since it can easily be generalized
for the higher dimensional cases.

Using the following entanglement strategy - inspired by the construction in (\ref{GHZ basis with Bell states}) - by applying Weyl transformations
\begin{eqnarray}
W_{k,l} =\sum_{s=0}^{d-1} w^{(s-l)k} |s-l\rangle \langle s|\qquad \mbox{with}\quad w=e^{\frac{2  \pi i}{d}}\quad\textrm{and}\quad k,l=0 \dots d-1,
\end{eqnarray}
for two particles of dimension $d=2$,  we obtain one of the Bell states
\begin{eqnarray}
|\Phi_2\rangle&=& \frac{1}{\sqrt{2}}\left( |0\rangle\otimes|\Phi_1\rangle+|1\rangle\otimes W_{1,1}|\Phi_1\rangle\right)\nonumber\\
&=& \frac{1}{\sqrt{2}}\left( |0\rangle\otimes |0\rangle+|1\rangle\otimes W_{1,1} |0\rangle\right)\nonumber\\
&=& \frac{1}{\sqrt{2}}\left( |0\rangle\otimes |0\rangle-|1\rangle\otimes|1\rangle\right)\nonumber\\
&=&|\phi^-\rangle\;.
\end{eqnarray}

In the next iteration a GHZ-type state is obtained
\begin{eqnarray}
|\Phi_3\rangle&=& \frac{1}{\sqrt{2}}\left( |0\rangle\otimes|\Phi_2\rangle+|1\rangle\otimes (\mathbbm{1}\otimes W_{1,1})|\Phi_2\rangle\right)\nonumber\\
&=&\frac{1}{\sqrt{2}}\left( |0\rangle\otimes|\phi^-\rangle-|1\rangle\otimes|\psi^+\rangle\right)\;.
\end{eqnarray}

Generally, for $n$ particles  of dimension $d=2$, we obtain  by entangled entanglement
\begin{eqnarray}
|\Phi_n\rangle&=& \frac{1}{\sqrt{2}}\left( |0\rangle\otimes|\Phi_{n-1}\rangle+|1\rangle\otimes (\underbrace{\mathbbm{1}\otimes \dots \otimes \mathbbm{1}}_{\substack{n-2}} \otimes W_{1,1}) |\Phi_{n-1}\rangle\right)\\
&=& \frac{1}{\sqrt{2}} \sum_{i=0}^{1} \left(\mathbbm{1}^{\otimes (n-1)}\otimes W_{i,i}\right)\; |i\rangle\otimes|\Phi_{n-1}\rangle\;.
\end{eqnarray}

A generalization for higher dimensions $d$ is possible in a similar way

\begin{eqnarray}
|\Phi_n^d\rangle&=& \frac{1}{\sqrt{d}}\left( |0\rangle\otimes|\Phi_{n-1}\rangle+|1\rangle\otimes (\underbrace{\mathbbm{1}\otimes \dots \otimes \mathbbm{1}}_{\substack{n-2}} \otimes W_{1,1}) |\Phi_{n-1}\rangle \dots +|d-1\rangle\otimes (\underbrace{\mathbbm{1}\otimes \dots \otimes \mathbbm{1}}_{\substack{n-2}} \otimes W_{d-1,d-1}) |\Phi_{n-1}\rangle\right)\nonumber\\
&=&\frac{1}{\sqrt{d}}\sum_{i=0}^{d-1} \left(\mathbbm{1}^{\otimes (n-1)} \otimes W_{d-1,d-1}\right)\;|i\rangle\otimes |\Phi_{n-1}\rangle\;.
\end{eqnarray}

Having obtained an entangled entanglement state for dimension $d$ and $n$ particles, we construct -- in every case -- the $d^n - 1$ remaining entangled entangled states
by acting in one of the subsystems with all $d^2$ Weyl operators and in $n-2$ subsystems with Weyl-operators  $W_{s,0}$ changing the phase, i.e.

\begin{eqnarray}\label{generaldefinitionstates}
|\Phi_n^d (s_1,s_2,\dots,s_{n-2},k,l)\rangle&=&\mathbbm{1}\otimes W_{s_1,0}\otimes \dots W_{s_{n-2},0}\otimes W_{k,l}|\Phi_{n}^d\rangle
\end{eqnarray}

\noindent and herewith we can construct our entangled entanglement simplex $\mathcal{W}^d_{n}$ for $n$-particles with dimension $d$:

\begin{eqnarray}
\mathcal{W}^d_{n}&:=&\{ \sum_{s_1,\dots,s_{n-2},k,l=0}^{d-1} c_{s_1,\dots,s_{n-2},k,l}|\Phi_n^d (s_1,s_2,\dots,s_{n-2},k,l)\rangle\langle \Phi_n^d (s_1,s_2,\dots,s_{n-2},k,l)|\nonumber\\
&&\qquad \textrm{with}\quad c_{s_1,\dots,s_{n-2},k,l}\geq 0\quad\textrm{and}\quad \sum c_{s_1,\dots,s_{n-2},k,l}=1\}\;.
\end{eqnarray}

Due to our construction the states can be written as:
\begin{eqnarray}\label{Summationterms}
|\Phi_n^d (s_1,s_2,\dots,s_{n-2},k,l)\rangle=&\frac{1}{\sqrt{d^{n-1}}} (&\gamma (0, \dots, 0, 0) |0 \dots 0 0 z (0, \dots, 0, 0)\rangle+\nonumber\\
&&\gamma (0, \dots, 0, 1) |0 \dots 0 1 z (0, \dots, 0, 1)\rangle \dots +\nonumber\\
&&\gamma (0, \dots, 0, d-1)|0 \dots 0(d-1) z (0, \dots, 0, d-1)\rangle +\nonumber\\
&&\dots  \dots\nonumber\\
&& \gamma (d-1, \dots, d-1, 0) |(d-1) \dots (d-1)0 z (d-1, \dots, d-1, 0)\rangle +\nonumber\\
&&\dots  \dots\\
&& \gamma (d-1, \dots, d-1, d-1)|\underbrace{(d-1),\dots,(d-1)}_{(n-1)-times}z (d-1, \dots, d-1, d-1)\rangle )\;,\nonumber
\end{eqnarray}
where the $\gamma(i_1,\dots,i_{n-1})$ are obtained from the different Weyl transformations, meaning that the coefficients are powers of the Weyl factor
$w^{\frac{2\pi i}{d}}$ and the $z(i_1,\dots, i_{n-1})$ are the results of the Weyl transformations $W_{k,l}$ for the last particle. The $z(i_1,\dots, i_{n-1})$ take different values, but all in all there are equally many results giving the "digits" $0,1,\dots, (d-1)$.

For all dimensions $d$ and particle numbers $n$, the $d^n$ states
\begin{eqnarray}
\{ |\Phi_n^d (s_1,s_2,\dots,s_{n-2},k,l)\rangle, 0\le s_1,s_2,\dots,s_{n-2},k,l \le d-1 \}
\end{eqnarray}
form an orthogonal system
\begin{eqnarray}
\langle|\Phi_n^d (s'_1,s'_2,\dots,s'_{n-2},k',l') |\Phi_n^d (s_1,s_2,\dots,s_{n-2},k,l)\rangle=\delta_{s'_1,s_1} \dots \delta_{s'_{n-2},s_{n-2}}\delta_{k',k}\delta_{l',l}\;.
\end{eqnarray}

\noindent\textbf{Proof:}\ \ \
\begin{eqnarray}
&&\langle \Phi_n^d (s'_1,s'_2,\dots,s'_{n-2},k',l')|\Phi_n^d (s_1,s_2,\dots,s_{n-2},k,l)\rangle\;=\nonumber\\
&&\quad=\;\langle \mathbbm{1}\otimes W_{s'_1,0}\otimes \dots W_{s'_{n-2},0}\otimes W_{k',l'}\, \Phi_n^d (0,\dots,0) |  \mathbbm{1}\otimes W_{s_1,0}\otimes \dots W_{s_{n-2},0}\otimes W_{k,l}\, \Phi_n^d  (0,\dots,0)\rangle\nonumber\\
&&\quad=\;\langle \Phi_n^d (0,\dots,0) |   \mathbbm{1} \otimes (W_{s'_1,0}^\dagger \cdot W_{s_1,0})\otimes  \dots \otimes (W_{s'_{n-2},0}^\dagger  \cdot W_{s_{n-2},0}) \otimes (W_{k',l'}^\dagger  \cdot W_{k,l}) |\Phi_n^d  (0,\dots,0)\rangle\nonumber\\
&&\quad=\;\langle \Phi_n^d (0,\dots,0) |   \mathbbm{1} \otimes  (W_{s'_1-s_1,0})\otimes  \dots \otimes  (W_{s'_{n-2 }-s_{n-2},0}) \otimes (w^{(k'-k)l'}  W_{k'-k,l'-l}) \Phi_n^d  (0,\dots,0)\rangle\;.
\end{eqnarray}
If $k \neq k'$ the Weyl transformation creates a different state moving $|s\rangle$ to $|s-(k-k')\rangle$ and therefore the inner product is equal to zero. If on the other hand $s_i \neq s_i'$ or $l \neq l'$, since $W_{j-j',0}$ is a diagonal matrix, the state does not change, but receives a new weight $w^{\frac{2\pi i}{d}.(j-j').s}$ with $s=s_i- s_i' \neq 0$.
We need to sum up all matching terms from (\ref{Summationterms}), terms showing the same "digit" respectively at the $i$-th place and collect the new weights.
We therefore obtain for this sum if $s_i \neq s_i'$
\begin{eqnarray}
d \cdot \sum_{j=0}^{d-1} w^{\frac{2\pi i}{d}\cdot j\cdot s} = 0\;.
\end{eqnarray}

\noindent Therefore the   $d^n$ states  form an orthogonal system. QED
\medskip

\begin{figure}[h!]
	\centering
	\includegraphics[width=0.7\textwidth]{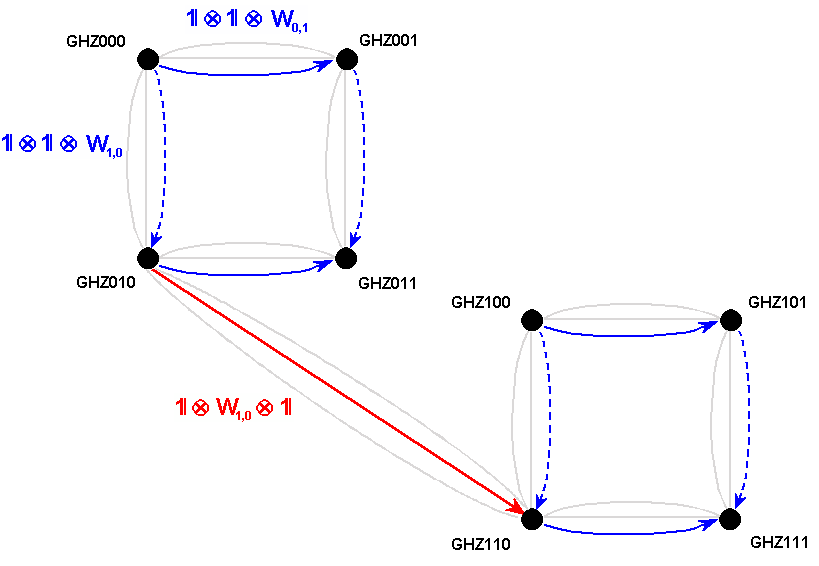}
	\caption{(Color online) Geometry of the basis states forming the magic simplex $\mathbbm{S}$ for three qubits: Here we simplify the notation by $|\Phi_3^2(s,k,l)\rangle=GHZ_3^2(s,k,l)=:GHZskl$. To change a certain GHZ state to another one within the square one needs to apply either a flip or phase operation in the last subsystem, whereas the phase operation $W_{1,0}$ applied to the second subsystem moves a certain GHZ state to a GHZ state in the other square.}
	\label{fig:GHZ states}
\end{figure}

\begin{figure}[h!]
	\centering
	\includegraphics[width=0.7\textwidth]{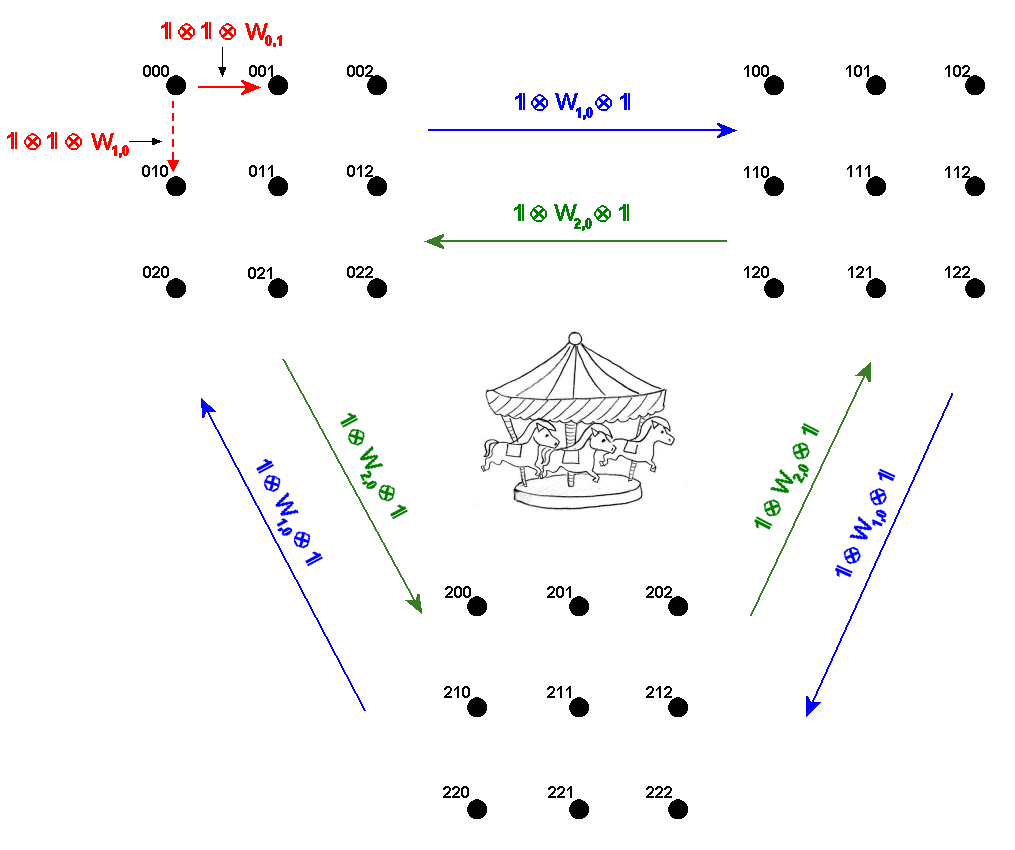}
	\caption{(Color online) \emph{``Sch\"on ist so ein Ringelspiel ...''}~\cite{Ringelspiel-Lied} or the \emph{Merry Go Round} of the GHZ states. In our construction the GHZ states possess a cyclic property that allows to move from square to the next one like in the Carousel of the famous Viennese Prater~\cite{ViennesePrater}. The geometry for three qutrits is a generalization of the one for three qubits, i.e. Fig.~\ref{fig:GHZ states}. Here we simplify the notation by $|\Phi_3^3(s,k,l)\rangle=GHZ_3^3(s,k,l)=:skl$. To change a certain GHZ state to another one within the square one needs to apply either a flip or phase operation in the third subsystem, whereas the phase operation $W_{1,0}$ or $W_{2,0}$ applied to the second subsystem moves a certain GHZ state form one square to a GHZ state in another square, yielding in this way the cyclic property in our construction of GHZ states. This can be extended straightforwardly to higher dimensions.}
	\label{fig:Qutrits}
\end{figure}

%
%

This general construction works for any number of particles $n$ and any dimension $d$ and reveals a particular geometry of the orthogonal basis of the GHZ-type of states. In Fig.~\ref{fig:GHZ states} we have depicted the geometry of three qubits. We find squares which are obtained by applying the four different Weyl operators $W_{0,0},W_{0,1},W_{1,0},W_{1,1}$ to one subsystem (in our choice to the last subsystem) of one reference GHZ-type state (denoted by $GHZ000$). The two squares are connected via applying one Weyl operator to another subsystem. This geometric structure generalizes for higher dimensions in an obvious way, illustrated for three qutrits in Fig.~\ref{fig:Qutrits}. We find three squares with each nine GHZ-type of states related by applying the nine Weyl operators to the last subsystem. To move from one square to another square one has to apply the phase shift operators in the second subsystem (as in the qubit case). Our construction shows that this straightforwardly generalizes for higher dimensions $d$ and more particles $n$.

Having defined state spaces via the convex combination of these orthogonal bases with the above described geometry, our magic simplexes, we proceed by discussing firstly the two theorems introduced previously by interferometric tools. Secondly, the geometry of entanglement including genuine multi-partite entanglement of states within these simplexes is analyzed. An introduction to partial entanglement is given before. Let us remark here also that the generality of our construction and the definition of the simplex allows in a defined and conceptually clear way to compare different number of particles $n$ and dimensions $d$.


\section{Interferometric Analyzes of the Transformation from Separable States into Entangled States}\label{trafoSepEntangle}

In this section we want to explore the information theoretic content of Theorem $1$ and Theorem $2$. We review first a general two-state interferometer and how Bohr's complementarity can be quantified, then we apply these considerations to our entangled entangling unitary operators.

\subsection{Interferometric Analyzes of Two-State Interfering Quantum Systems}

Let us consider any two-state interfering system and analyze its information theoretic content~\cite{Englert,bramon}. E.g. think about an interferometer, i.e. a beam splitted and afterwards merged again, inbetween a phase shift $\phi$ between the two paths is introduced. Prior to entering the interferometer our two level system is given by
\begin{eqnarray}
\rho_0&=&\frac{1}{2}\{\mathbbm{1}+\vv{n}\vv\sigma\}\;,
\end{eqnarray}
where the Blochvector is given by $\vv{n}=Tr\{\vv\sigma\rho_0\}$ and the $\sigma$'s are the usual Pauli matrices.

\begin{figure}
	\centering
	\includegraphics[width=0.8\textwidth]{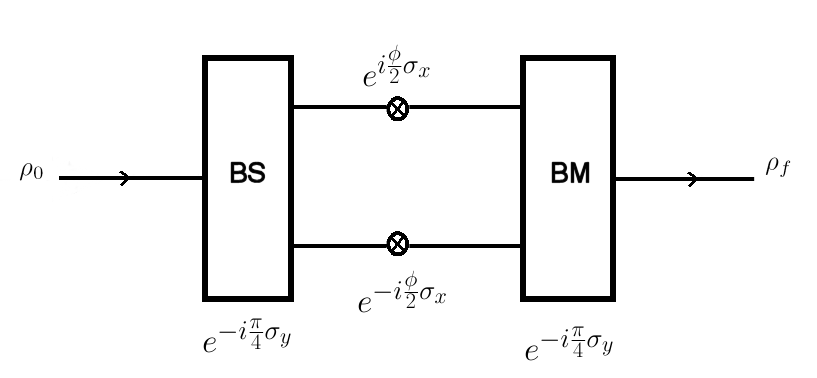}
	\caption{The picture shows the very working of an interferometric device. An initial state $\rho_0$ traverses a beam splitter (BS), in each path picks up an phase $\frac{\phi}{2}$ and after the beam merger (BM) a final state $\rho_f$ is obtained. Bohr's complementarity relation states the total information content stored in $\rho_0$ is not changed, however, depending on initial state different a priori fringe visibilities are observed.}
	\label{fig:Interferometer}
\end{figure}

The effect of the beam splitter (BS) and the beam merger (BM) can be described reasonably well by the operator
\begin{eqnarray}
U_{BS}=U_{BM}=e^{-i \frac{\pi}{4} \sigma_y}\;,
\end{eqnarray}
whereas the phase difference can be introduced in direction $x$ by
\begin{eqnarray}
U_{phase}=e^{-i \frac{\phi}{2} \sigma_x}\;,
\end{eqnarray}
see Fig.~\ref{fig:Interferometer}. Then the final state after the beam merger is given by
\begin{eqnarray}
\rho_f=e^{-i \frac{\pi}{4} \sigma_y}e^{-i \frac{\phi}{2} \sigma_x}e^{-i \frac{\pi}{4} \sigma_y}\;\rho_0\;e^{i \frac{\pi}{4} \sigma_y}e^{i \frac{\phi}{2} \sigma_x}e^{i \frac{\pi}{4} \sigma_y}\;,
\end{eqnarray}
which gives
\begin{eqnarray}
\rho_f=\frac{1}{2}\{\mathbbm{1}+\left(\begin{array}{c}-n_x \cos\phi+n_y \sin\phi\\
n_x \sin\phi+n_y \cos\phi\\-n_z\end{array}\right)\vv\sigma\}\;.
\end{eqnarray}

From that we obtain the probability to obtain $+$ or $-$ in $z$-direction, i.e. in the computational basis, by
\begin{eqnarray}
Prob(\pm z)=Tr(\frac{1}{2}(\mathbbm{1}\pm\sigma_z)\rho_f)=\frac{1}{2}(1\mp|n_z|)
\end{eqnarray}
and therefore the predictability can be defined by
\begin{eqnarray}
\mathcal{P}:=|Prob(+)-Prob(-)|=|n_z|\;.
\end{eqnarray}
Note that in case the second beam splitter acting as a merger is not applied, this is equivalent to the above case! A measurement in $x$- and $y$-direction gives
\begin{eqnarray}
Prob(\pm x)&=&Tr(\frac{1}{2}(\mathbbm{1}\pm\sigma_x)\rho_f)=\frac{1}{2}(1\pm(-n_x \cos\phi+n_y \sin\phi))\equiv\frac{1}{2}(1\pm\sqrt{(n_x)^2+(n_y)^2}\cos\Phi)\nonumber\\
Prob(\pm y)&=&Tr(\frac{1}{2}(\mathbbm{1}\pm\sigma_y)\rho_f)=\frac{1}{2}(1\pm(n_x \sin\phi+n_y \cos\phi)\equiv\frac{1}{2}(1\pm\sqrt{(n_x)^2+(n_y)^2}\sin\Phi)
\end{eqnarray}
which defines the \textit{a priori} fringe visibility
\begin{eqnarray}
\mathcal{V}_0&=&\sqrt{(n_x)^2+(n_y)^2}\;,
\end{eqnarray}
i.e. the fringe contrast (difference between maximum and minimum in the interference pattern).

Obviously, we have
\begin{eqnarray}
\mathcal{P}^2+\mathcal{V}_0^2=(n_z)^2+(n_x)^2+(n_y)^2=Tr(\rho_0^2)\leq 1\;.
\end{eqnarray}
Equality holds for pure states ($Tr(\rho_0^2)=1$). This is Bohr's famous complementarity relation in a quantitative formulation,  gaining more which way information (``particle property'') reduces the interference contrast (``wave property''). For example, in Ref.~\cite{CPBohr} the authors showed that the tiny difference between a world and an antiworld, more precise the non-conservation of the parity--charge-conjugation symmetry, observed in the neutral K-meson systems shifts obtainable information about our reality, visibility and predictability, accordingly to Bohr's complementarity principle.

\subsection{Interferometric Analyzes of the Transformation from Separable States into GHZ States}\label{trafoSepGHZ}

Before we analyze the information theoretic content analogously to the interferometric analyses given in the previous section we introduce a useful parametrization of unitary operators.

For any unitary operation $U$ acting on a Hilbert space $\mathcal{H}=\mathbb{C}^d$ with $d\geq2$ spanned by the orthonormal basis $\{ \ket{1},\ldots,\ket{d} \}$ there exist $d^2$ real values $\lambda_{m,n}$ with $m,n \in \left\{ 1 , \ldots , d \right\}$ and $\lambda_{m,n} \in \left[0, 2 \pi \right]$ for $m \geq n$ and $\lambda_{m,n} \in \left[0, \frac{\pi}{2} \right]$ for $m < n$ such that any $U\equiv U_C$ with (this concise and intuitive parameterization was introduced in Ref.~\cite{SHH1}; it allows for example to derive analytically the Haar measure for unitary groups~\cite{SHH2})
\beq
\label{Uc}
U_C=\left[\prod_{m=1}^{d-1} \left(\prod_{n=m+1}^{d} \mbox{exp} \left( i\, {\color{blue}{\lambda_{n,m}}}\;P_n  \right) \mbox{exp} \left( i\, {\color{green}{\lambda_{m,n}}}\;\sigma_{m,n} \right)  \right) \right] \cdot \left[ \prod_{l=1}^{d} \mbox{exp}(i\, {\color{red}{\lambda_{l,l}}}\;P_l )\right] \ .
\eeq
The sequence of the product is defined by $\prod_{i=1}^{N}A_i=A_1 \cdot A_{2} \cdots A_N$.
Here, the $P_l $ are one-dimensional projectors
\beq
P_l=\ket{l}\bra{l}
\eeq
and $\sigma_{m,n}$ are the generalized anti-symmetric $\sigma$-matrices
\beq
\sigma_{m,n}=-i\ket{m} \bra{n} + i \ket{n} \bra{m}
\eeq
with $1\leq m < n \leq d$.\\

The parameter $\lambda_{m,n}$ can be gathered in a $d \times d$ ``\textit{parameterization matrix}''
\beq
\textrm{``\textit{Parameterization matrix}''}\;\equiv\;\left(
  \begin{array}{ccccc}
    {\color{red}{\lambda_{1,1}}} &{\color{green}{\lambda_{1,2}}}&   \cdots &{\color{green}{\lambda_{1,d-1}}}& {\color{green}{\lambda_{1,d}}} \\
    {\color{blue}{\lambda_{2,1}}} &{\color{red}{\lambda_{2,2}}}&   \cdots &{\color{green}{\lambda_{2,d-1}}}&{\color{green}{\lambda_{1,d}}} \\
    \vdots &\vdots & \ddots & \vdots \\
   {\color{blue}{\lambda_{d-1,1}}} &{\color{blue}{\lambda_{d-1,2}}}&   \cdots &{\color{red}{\lambda_{d-1,d-1}}}& {\color{green}{\lambda_{d-1,d}}} \\
    {\color{blue}{\lambda_{d,1}}} &{\color{blue}{\lambda_{d,2}}}&   \cdots &{\color{blue}{\lambda_{d,d-1}}}& {\color{red}{\lambda_{d,d}}}
  \end{array}
\right) \ ,
\eeq
where the diagonal entries ${\color{red}{\lambda_{n,n}}}$ represent global phase transformations, the upper right entries ${\color{green}{\lambda_{m,n}}}$ are related to rotations in the subspaces spanned by $|n\rangle$ and $|m\rangle$, while the lower left entries ${\color{blue}{\lambda_{n,m}}}$ are relative phases in these subspaces (with respect to the basis $\{\ket{1},\ldots,\ket{d}\}$).

Let us start with theorem $1$ and the initial separable states $|RR\rangle$ or $|LL\rangle$. We apply to both subsystems a beam splitter operation and then a phase shift of $\frac{\pi}{2}$ in each of the four interfering paths is introduced. The phase shift $\frac{\pi}{2}$ is chosen since it transforms $|R\rangle/|L\rangle$ solely into $|H\rangle/|V\rangle$ contributions. Then we apply the entangling unitary operator
\begin{eqnarray}
 U_{ent}=e^{i  \frac{\pi}{4} \sigma_{1,4}}=e^{i  \frac{\pi}{4}(-i |00\rangle\langle 11|+i|11\rangle\langle 00|)}
\end{eqnarray}
onto the four interfering paths where only two should be nonzero. Next, applying the beam merger we turned the separable states $|RR\rangle$ or $|LL\rangle$ into the Bell states in the computational basis $-|\phi^-\rangle$, $i |\psi^+\rangle$, whereas choosing $|RL\rangle$ or $|LR\rangle$ as initial states leads to separable output states. Choosing as initial states the basis states of the computational basis $|00\rangle,|11\rangle,|01\rangle,|10\rangle$ leads to non-maximally entangled states. Indeed, the concurrence $\mathcal{C}$, a measure of bipartite entanglement, equals $\frac{1}{2}$. The same holds true if we choose as initial states the eigenstates of the $\sigma_x$ operator, i.e. $|\pm 45^\circ\rangle=\frac{1}{\sqrt{2}}(|0\rangle\pm|1\rangle)$. Consequently, we observe that the total entanglement, i.e. $\mathcal{C}=2$, is conserved by the unitary gate. Again the choice of the basis of the involved operators and states makes the physical difference of the revealed information content, Bohr's complementarity is at work! In our case the $\{R,L\}$ basis concentrates entanglement to only two states!

\begin{figure}
	\centering
	\includegraphics[width=1\textwidth]{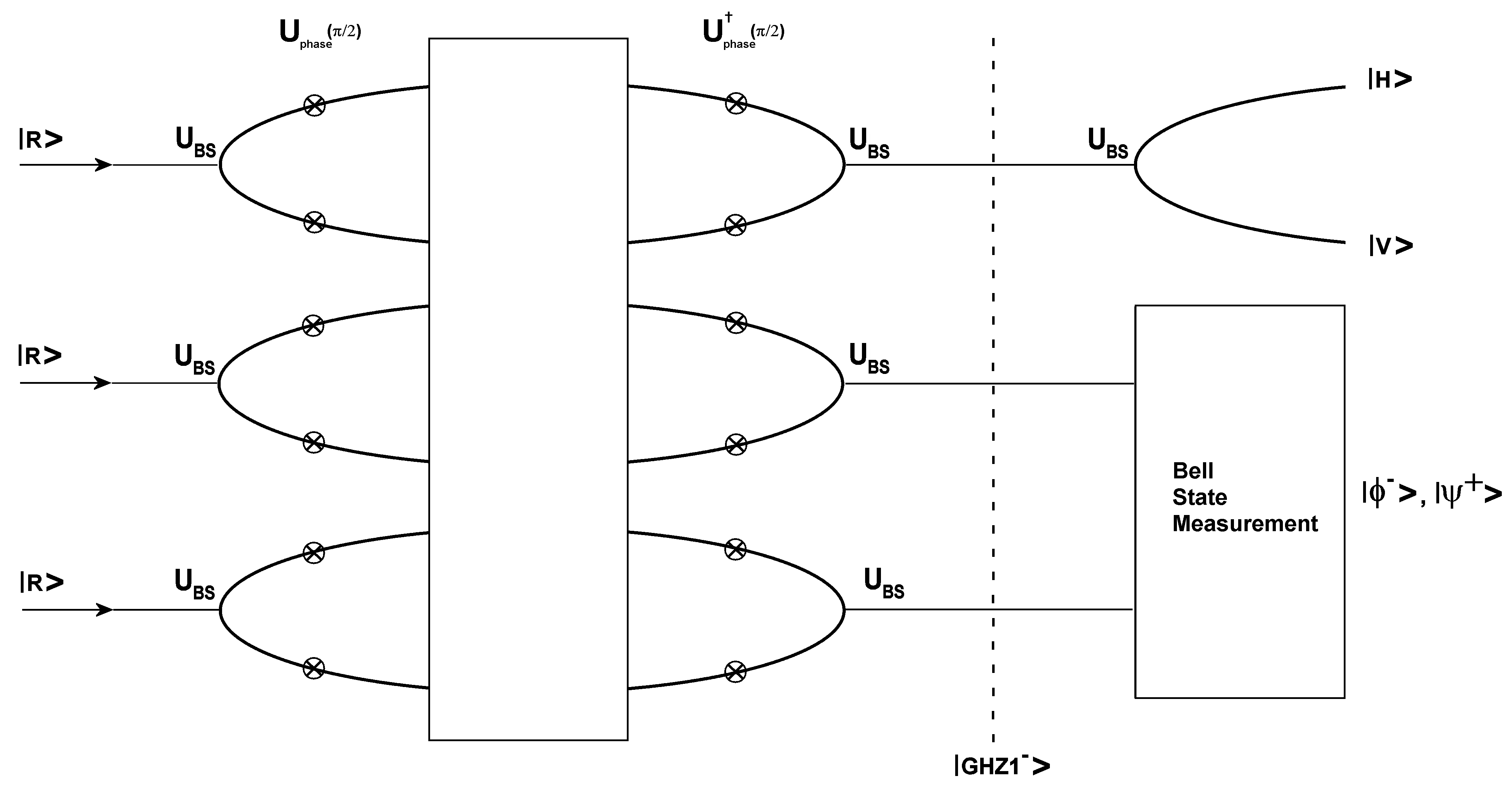}
	\caption{GHZ-Interferometer: This schematic illustration shows how product states transform into maximally entangled $GHZ$ states and how the entangled entanglement structure can be revealed by suitable measurements. If the initial states are given in the basis $\{R,L\}$ and all either $R$ or $L$ then the final state after the GHZ-interferometer is a GHZ state. Applying a third beam splitter $U_{BS}$ to one subsystems (or equivalently removing one beam splitter in one path) entangles the other two pathes. (Compare with Theorem 1 and Theorem 2.)}
	\label{fig:GHZ-Interferometer}
\end{figure}

The same procedure extends to three particles (theorem 2), the entangling operator in this case is given straightforwardly by
\beq
U_{ent}=e^{i  \frac{\pi}{4} \sigma_{1,8}}=e^{i  \frac{\pi}{4}(-i |000\rangle\langle 111|+i|111\rangle\langle 000|)}\;.
\eeq
In Fig.~\ref{fig:GHZ-Interferometer} we schematically draw the GHZ-interferometer. A measurement after the three beam splitter, phase shifter, entangling operator and then the three beam merger results in a three-particle entangled state, the GHZ-state. Removing one beam splitter or equivalently applying another beam splitter for one subsystem reveals the entangled entanglement structure. Compare also with Fig.~\ref{fig:Bob photons entangled}.

Any initial state in the $\{R,L\}$ basis that is not $|RRR\rangle$ or $|LLL\rangle$ leads to a separable state.  Choosing as initial states any of the eight basis states in the computational basis $\{H,V\}$ or in the basis $\{+45^\circ,-45^\circ\}$ results in final states with equal entanglement content of each output state, however, which is not maximally entangled (e.g., detected via the later introduced criterion $Q_{GHZ}$ of the HMGH-framework (Huber, Mintert, Gabriel, Hiesmayr)~\cite{hmgh}, Eq.~(\ref{criteriondetailed}); only GHZ-states account for the maximal value $Q_{GHZ}=1$ whereas the results for the basis $\{H,V\}$ equals to $Q_{GHZ}=0.432541$ for each basis state and for the basis $\{+45^\circ,-45^\circ\}$ equals to $Q_{GHZ}=0.325194$ for each basis state). Generally, we find that the above defined GHZ-interferometer concentrates only in one combination of three mutually bases genuine multi-partite entanglement whereas for all other basis choices entanglement is equally distributed among the output states, however, the absolute values of entanglement are different and not maximal. We can, of course, also consider input states of different basis choices, then we find separable and non-maximally entangled output states. Summing up, maximal genuine multi-partite entanglement is only generated if maximal visibility of each subsystem by the setup is guaranteed!

%
%
%


%
%
%


\section{The Geometry of Partial Separability \& Genuine Multi-partite Entanglement in the Magic Simplex}\label{sec:separability multipartite entanglement}

In Refs.~\cite{hmgh,ghh}, a universal framework for constructing very general separability criteria for quantum states living in finite and continuous Hilbert spaces~\cite{hmghcontinous} and for arbitrary number of particles was introduced, dubbed the HMGH-framework~\cite{hmgh}. We want here to give a pedagogical introduction into partial separability, genuine multi-partite entanglement and the HMGH-framework. In a next step we apply this HMGH-framework to our entangled entanglement classes of states. Herewith we obtain the geometric properties of different types of entangled entanglement. Moreover, our construction leading to the  magic simplex allows us in a well defined way to compare different dimensions and number of particles.

\subsection{Definition of Partial Separability \& Genuine Multi-partite Entanglement}

As is well-known the problem of determining whether an arbitrary quantum state is entangled or not (where in
the latter case, the state is called separable), dubbed the Quantum Separability Problem,
was proven to be NP-hard by Gurvits in 2003~\cite{Gurvits}($NP$\dots $N$on-deterministic, $P$olynomial-time hard problem in computational complexity theory; it is a class of problems that are, loosely spoken, at least as hard as the hardest problems in this set). This means that all existing entanglement criteria solve the problem of detecting and characterizing entanglement only for certain classes of states and, in addition, some of them are very resource intensive if applied experimentally. For example, to apply the well-known Peres-Horodecki
criterion~\cite{peres, horodecki-M-P-R-M96}, i.e. operating with the partial transpose transformation on one subsystem, requires obviously the full information on the state, i.e. a full quantum state tomography in practice. The criteria of the HMGH-framework are formulated as inequalities for mean values of few observables and therefore they are especially useful to detect entanglement in experiments. Moreover, as we will show one can formulate them to detect different types of multi-partite entanglement.

For example, the HMGH-framework allowed the first experimental proof that single neutrons in an interferometric setup can be produced in a genuine multi-partite entangled state~\cite{neutron}, including the GHZ-type of genuine multi-partite entanglement. Here, the outer degrees of freedom, the path in the interferometer, and the inner degrees of freedom, i.e. spin and energy eigenstates, were considered. Another example concerns quantum cryptography. With the help of this HMGH-framework a simple proof is obtained to show how the special type of GHZ-state entanglement secures secret sharing protocols~\cite{SHH}. The main idea is to divide a secret into several shares and distribute these shares among several parties such that each party alone is not able to gain any information about the secret.

For bipartite entangled systems the quantum separability problem reduces to the question whether the state is entangled or not, however, in the multi-partite case the problem is more involved. First of all, there exist different hierarchies of entanglement since an $n$-partite entangled state $\rho$ may be a convex combination of pure entangled states with maximally $k$ entangled particles. For example, any tripartite pure qudit state can be written by
\beq
|\psi_{k=3}\rangle&=&|\phi_A\rangle\otimes|\phi_B\rangle\otimes|\phi_C\rangle\nonumber\\
|\psi_{k=2}\rangle&=&|\phi_A\rangle\otimes|\phi_{BC}\rangle,\quad|\phi_B\rangle\otimes|\phi_{AC}\rangle\quad\textrm{or}\quad |\phi_{AB}\rangle\otimes|\phi_{C}\rangle\nonumber\\
|\psi_{k=1}\rangle &=&|\psi\rangle_{ABC}
\eeq
where $k$ gives the number of partitions corresponding to the $k$-separability that we defined in detail later (see Eq.~(\ref{k-separability})). Here, we find already another interesting complication of the factorization since for $k=1$ the state can be physically entangled in different ways, i.e.
\beq
|GHZ\rangle&=&\frac{1}{\sqrt{2}}\{|000\rangle+|111\rangle\}\nonumber\\
|W\rangle&=&\frac{1}{\sqrt{3}}\{|001\rangle+|010\rangle+|100\rangle\}\;.
\eeq
Certainly, if Alice finds her particle to be in the state $0$ in the case of the $GHZ$ state Bob's and Charlie's particle are in a separable state, however, in case of the $W$ state we find that Bob and Charlie share a maximally entangled state. Therefore, we would like to have in general a finer classification and detection tools to distinguish different types of $k$-separable states.

 Let us therefore exactly define $k$-separability. A pure state $\ket{\Psi^k}$ is called $k$-separable, if and only if ($:=$ iff) it can be written as a tensor product of $k$ factors $\ket{\psi_i}$, each of which describes one or several subsystems:
\beq\label{k-separability}
\ket{\Psi^k}\;=\;|\psi_1\rangle\otimes\ket{\psi_2}\otimes\dots\otimes\ket{\psi_k}\;=\;|\psi_1\psi_2\dots\psi_k\rangle\;=\; \bigotimes_{i=1}^k \ket{\psi_i} \; .\eeq

A mixed state $\rho$ is called $k$-separable, iff it can be decomposed into a mixture of $k$-separable pure states:
\beq \rho = \sum_i p_i \ket{\Psi_i^k}\bra{\Psi_i^k} \eeq
where all $\ket{\Psi_i^k}$ are $k$-separable (possibly with respect to different $k$-partitions) and the $p_i$ form a probability distribution. An $n$-partite state (pure or mixed) is called fully separable iff it is $n$-separable. It is called genuinely multi-partite entangled (GME) iff it is not biseparable (2-separable). If neither of these is the case, the state is called partially entangled or partially separable. Note that obviously a $k=3$ separable state is necessarily also $k=2$ separable, thus $k$-separable states have a nested-convex structure.

In particular, note that the following tripartite mixed state
\beq
\rho&=&\sum_i p_i\; |\psi_i\rangle_{AB}\langle\psi_i|_{AB}\otimes|\sigma_i\rangle_C\langle \sigma_i|_C+\sum_i q_i\; |\chi_i\rangle_{AC}\langle\chi_i|_{AC}\otimes|\tau_i\rangle_B\langle \tau_i|_B+\sum_i r_i\; |\xi_i\rangle_{BC}\langle\xi_i|_{BC}\otimes|\omega_i\rangle_A\langle \omega_i|_A
\eeq
with $p_i,q_i,r_i\geq 0$ and $\sum p_i+q_i+r_i=1$ is biseparable though it is not biseparable with respect to a certain splitting. This property and the fact that the convex sum of pure states is not unique are the reasons why it is so hard to detect genuine multi-partite entanglement, i.e. a state that cannot be written in the above form. Consequently, the entanglement characterization of multi-partite states needs more than the combination of bipartite entanglement criteria.

\subsection{The HMGH-Framework to Detect $k$-Separability}

Let us start the development of our criteria by considering pure states $\psi$ and the Cauchy-Schwarz inequality
\beq
\left|x\cdot y\right|\leq\left|x\right|\cdot \left|y\right|\;=\; \sqrt{\left|x\right|^2\cdot \left|y\right|^2},
\eeq
where we set the complex numbers equal to $x=\langle \chi_1|\psi\rangle$ and $y=\langle \chi_2|\psi\rangle$ such that the Cauchy-Schwarz inequality re-writes to
\beq\label{Cauchy}
|\langle \chi_1|\rho|\chi_2\rangle|\;\leq\;\sqrt{\langle \chi_1|\otimes \langle\chi_2|\;\rho\otimes\rho\;|\chi_1\rangle\otimes|\chi_2\rangle}\;=\;\sqrt{\langle \chi_1 \chi_2|\rho^{\otimes 2}|\chi_1\chi_2\rangle},
\eeq
where we used $\rho=|\psi\rangle\langle\psi|$. Further we want to define permutation operators acting on $\rho\otimes\rho$ such that a certain partition $\alpha$ of $\rho$ is interchanged by its copy, e.g.
\beq
P_{\alpha=\{1,2\}} |a_1 a_2 a_3\cdots a_n\rangle\otimes|a'_1 a'_2 a'_3\cdots a'_n\rangle&=& |a'_1 a'_2 a_3\cdots a_n\rangle\otimes|a_1 a_2 a'_3\cdots a'_n\rangle\;.
\eeq

Now let us assume that the pure state is separable concerning a specific partition $\alpha=\{12|345\}$, e.g., we choose the following five-partite state
\beq
|\psi^{12|345}\rangle=|\psi_{12}\rangle\otimes|\psi_{345}\rangle\;.\eeq
Obviously, applying the permutation operator $P_{\{1,2\}}$ to $\rho\otimes\rho$ has to be a symmetry, i.e. the commutator vanishes
\beq
\left[P_{\{1,2\}},\rho^{12|345}\otimes\rho^{12|345}\right]=0\qquad\Longleftrightarrow\qquad (\rho^{12|345})^{\otimes2}=P_{\{1,2\}}^\dagger\; (\rho^{12|345})^{\otimes2}\; P_{\{1,2\}}\;,
\eeq
since the state under consideration is separable with respect to this permutation of the first and second particle between the original and copy state. Using this symmetry by replacing $\rho^{\otimes 2}$ by $P_{\alpha=\{1,2\}}^\dagger\;\rho^{\otimes 2}\; P_{\alpha=\{1,2\}}$ in the Cauchy-Schwarz inequality~(\ref{Cauchy}), we obtain a necessary condition that has to hold for any state that is separable with respect to the chosen partition $\{12|345\}$. This means that if we find a violation of the inequality for any given state, we know it is not separable with respect to this chosen partition!

Due to the convexity property of the Cauchy inequality this inequality holds also for all mixed states $\rho$. Obviously, it is a necessary but not sufficient criterion, however, as shown for various quantum systems of different dimension and number of particles it turns out to be a surprisingly powerful criterion (see e.g. Refs.~\cite{c1,c2,c3,c4,c5,c6,c7,c8}) and in some cases also equivalent to the lower bound of measures of genuine multi-partite entanglement~\cite{mconcurrence}.

In case we want to have a criterion that is satisfied for any given $k$--separable state we simply have to sum over all permutation operators corresponding to all possible $k$-partitions. To obtain a concise formulation let us first introduce a total permutation operator $P_{total}$ which wholly permutes two arbitrary states $\ket{\chi_1},\ket{\chi_2}$, i.e.
\beq P_{total}\ket{\chi_1}\otimes\ket{\chi_2}&=&\ket{\chi_2}\otimes\ket{\chi_1}\;.\eeq With the help of this operator we can re-write the right hand side of the Cauchy inequality by
\beq
\langle \chi_1|\rho|\chi_2\rangle= \sqrt{\bra{\chi}\rho^{\otimes 2} P_{total}\ket{\chi}}\;.
\eeq
Let us assume again our state is biseparable under the specific partition $\alpha=\{\alpha_1|\alpha_2\}=\{12|345\}$, then the commutator with the $P_{\alpha_1=\{1,2\}}$ but also with $P_{\alpha_2=\{3,4,5\}}$ have to vanish. Thus for any such state we have
\beq\label{rhs}
\sqrt{\langle \chi|\rho^{\otimes 2}|\chi\rangle}&=&\sqrt{\left(\langle \chi|P^\dagger_{\alpha_1=\{1,2\}}\rho^{\otimes 2}P_{\alpha_1=\{1,2\}}|\chi\rangle\cdot\langle \chi|P^\dagger_{\alpha_2=\{3,4,5\}}\rho^{\otimes 2}P_{\alpha_2=\{3,4,5\}}|\chi\rangle\right)^{\frac{1}{2}}}\;.
\eeq
Straightforwardly we can extend the product for higher partial separability partitions $k$'s.

Consequently, for a given $k$ we have to subtract from the off-diagonal term of $\rho$ (given by equation~(\ref{rhs}) that follows from the r.h.s. of the Cauchy-Schwarz inequality~(\ref{Cauchy}))
 the sum over all
 possible $k$-partitions which are products of all possibilities of the specific permutation $\alpha_i$.  Thus the following criterion holds for all $k$-separable states $\rho$
\begin{equation}\label{k-separabilitycriterion} 
I_k:=\sqrt{\bra{\chi}\rho^{\otimes 2} P_{total}\ket{\chi}} - \sum_{\{\alpha\}}\left(\prod_{i=1}^k \bra{\chi}P_{\alpha_i}^\dagger \rho^{\otimes 2} P_{\alpha_i} \ket{\chi}\right)^{\frac{1}{2k}} \leq 0\;. \end{equation}

\subsection{The HMGH-Framework for Different Types of Genuine Multi-Partite Entanglement}

Our next goal is to distinguish between different types of genuine multi-partite entangled states, e.g., between the GHZ-type and the W-type of entanglement for $n$ qudits. Let us start with the GHZ-type of entanglement. The above inequality picks via $\ket{\chi}$ a certain off-diagonal element of the density matrix $\rho$ and the sum corresponds to diagonal elements of the density matrix $\rho$. Thus for a given pure GHZ-state in the computational basis the maximum violation of the criterion is given if we choose for $\ket{\chi_1}=|0\rangle^{\otimes n}$ and for $\ket{\chi_2}=|1\rangle^{\otimes n}$ since in this case we pick out the only non-zero off-diagonal element $\bra{0}^{\otimes n}\rho|1\rangle^{\otimes n}$, i.e. we obtain from Eq.(\ref{k-separabilitycriterion})
\beq Q_{GHZ}(\rho)=|\bra{0}^{\otimes n}\rho\ket{1}^{\otimes n}| - \sum_{\gamma}\sqrt{\bra{0}^{\otimes n}\bra{1}^{\otimes n}\mathcal{P}_{\gamma_A}^\dagger \rho^{\otimes 2} \mathcal{P}_{\gamma_A} \ket{0}^{\otimes n}\ket{1}^{\otimes n}}\;\leq 0\;, \eeq
which has to hold for all bi-separable states where the sum runs over all bi-partitions $\gamma=\{A,B\}$. The permutation operators $\mathcal{P}_{\gamma_A}$ permute the two copies of all subsystems contained in the first part of $\gamma$.

Let us consider as an example a general tripartite qubit state $\rho$, then we have the following permutations $\mathcal{P}_{\gamma_A}=\{12,3\},\{2,13\},\{1,23\}$ and thus we obtain (we introduce here an arbitrary factor $2$ in order to re-scale the function to be normalized for qubits)
\beq\label{criteriondetailed}
 Q_{GHZ}(\rho)&=&2\cdot\left(\sqrt{\langle 000|\otimes \langle 111| \rho^{\otimes 2} |111\rangle\otimes|000\rangle}-\sqrt{\langle 110|\otimes \langle 001|\rho^{\otimes 2}|110\rangle\otimes|001\rangle}-\sqrt{\langle 101|\otimes \langle 010|\rho^{\otimes 2}|101\rangle\otimes|010\rangle}\right.\nonumber\\
&&\left.-\sqrt{\langle 011|\otimes \langle 100|\rho^{\otimes 2}|011\rangle\otimes|100\rangle}\right)\nonumber\\
&=&2\cdot\left(|\langle 000|\rho|111\rangle|-\sqrt{\langle 110|\rho|110\rangle\langle 001|\rho|001\rangle}-\sqrt{\langle 101|\rho|101 \rangle\langle 010|\rho|010\rangle}-\sqrt{\langle 011|\rho|011 \rangle\langle 100|\rho|100\rangle}\right)\;.
\eeq
Obviously, when the state $\rho$ is a pure GHZ state given in the computational basis only the first term is nonzero. Since this off-diagonal term cannot be larger than given by the GHZ state this proves together with the fact that any non-zero diagonal element not equal $\langle 000|\rho|000\rangle$ or $\langle 111|\rho|111\rangle$ would only lower the value of the function $Q_{GHZ}$ that the maximum violation is obtained by the GHZ state.

Let us make some remarks here concerning the optimal violation of the criterion with respect to a given state $\rho$. There are mainly two conditions that optimize the criteria: (a) if $\langle \chi_1|\chi_2\rangle=0$ is chosen and (b) if $|\langle \chi_1|\rho|\chi_2\rangle|$ is chosen to be maximal since it is the only positive element.

 In other cases one may be interested to have a criterion that maximizes for other genuine multi-partite entangled states such as the $W$ state or generally for Dicke states. In general such criteria $Q_{Dicke}$ have been introduced in Ref.~\cite{DickeCriteria}. In particular the criteria $Q_{Dicke,n}^m$ assume their respective maximal values for the corresponding $n$-partite Dicke-state with $m$ excitations
\begin{eqnarray}\label{dicke}
|D_n^m\rangle&=& \frac{1}{\tiny{\sqrt{\left(\begin{array}{c}n\\m\end{array}\right)}}}\sum_{\{\beta\}}|d_\beta\rangle\;,
\end{eqnarray}
where the set of indices $\{\beta\}$ corresponds to the respective subsystems of excitations and the sum is taken over all inequivalent sets $\{\beta\}$ fulfilling $|\{\beta\}|=m$ and  $\ket{d_\beta}$ is the product state vector with $\ket{1}$ in all subsystems $i \in \beta$ and $\ket{0}$ otherwise, i.e.
 \begin{eqnarray}
 |d_\beta\rangle &=& \bigotimes_{i\not\in \beta} |0\rangle_i\bigotimes_{i\in\beta} |1\rangle_i\;.
 \end{eqnarray}
 In the case of $m=1$, these states are the $W$ states.

These criteria are defined by
\beq\label{Dicke} Q_{Dicke,n}^m(\rho) &=& \sum_{\sigma}\left(|\bra{d_\alpha}\rho\ket{d_\beta}| - \sqrt{\bra{d_\alpha}\bra{d_\beta} \mathcal{P}_\alpha^\dagger\rho^{\otimes 2} \mathcal{P}_\alpha \ket{d_\alpha}\ket{d_\beta}}\right)\nonumber\\
&& - m(n-m-1)\sum_\beta\bra{d_\beta}\rho\ket{d_\beta} \eeq
for $1 \leq m \leq \left\lfloor\frac{n}{2}\right\rfloor$, where the first sum runs over all sets $\sigma = \{\alpha,\beta\}$ with $\alpha,\beta \subset \{1,2,...,n\}$ such that $|\alpha|=|\beta|=m$ and $|\alpha \cap \beta| = (m-1)$.
The permutation operators $\mathcal{P}_\alpha$ permute all subsystems in $\alpha$ with their respective copies in the two-copies of $\rho$.

As an explicit example let us consider this criterion for a tripartite qubit state with one excitation
\beq\label{Dicketripartite}
 Q_{Dicke,3}^1(\rho)&=& 2 Re\{\langle 001|\rho|010\rangle\}+2 Re\{\langle 001|\rho|100\rangle\}+2 Re\{\langle 010|\rho|100\rangle\}\nonumber\\
 &&-\left(\langle 001|\rho|001\rangle+\langle 010|\rho|010\rangle+\langle 100|\rho|100\rangle\right.\nonumber\\
 &&\left.+ 2 \sqrt{\langle 000|\rho|000\rangle\cdot \langle 011|\rho|011\rangle}+2 \sqrt{\langle 000|\rho|000\rangle\cdot \langle 101|\rho|101\rangle}+2 \sqrt{\langle 000|\rho|000\rangle\cdot \langle 110|\rho|110\rangle}\right)\;.
\eeq
The positive terms are exactly the only non-zero off-diagonal terms of the $W$ state in the computational basis, while the negative terms are only diagonal terms. Again these negative terms are all zero for the $W$ state in the computational basis such that this state obtains the maximum value. Differently to the criterion optimized for GHZ-type of entanglement one has to subtract more terms from the diagonal elements of the density matrix. It is straightforward to see that the above Dicke-type criterion is zero for a GHZ-state given in the computational basis, however, optimization over the basis choices gives also a non-zero value for a general GHZ state. This basis dependence can be exploited to guarantee the security of secret sharing protocols as has been shown by the authors of Ref.~\cite{SHH}.

In general, one may be interested to investigate an a-priori detection probability of genuine multi-partite entanglement for these criteria if one does not have knowledge about the basis in which a state is produced by the source or how a channel decoheres it or about the very working of the detectors used. These questions were analyzed in Ref.~\cite{deviceindependGME} in details, where a measurement-device independence of the criteria based on the HMGH-framework was shown to be in most cases surprisingly high.

\subsection{The Geometry of Entangled Entanglement in the Magic Simplex}

In this section we want to discuss how the entanglement features, in particular the genuine multi-partite entanglement, changes when we mix certain GHZ states, i.e. the corner states of our simplex. We start  by showing how the geometry of the genuine multi-partite entanglement is structured within the magic simplex of entangled entanglement for three particles and different dimensions and proceed then to higher number of particles.
\begin{figure}
\flushleft{(a)\includegraphics[width=0.3\textwidth]{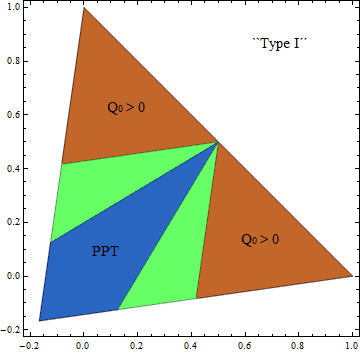}(b)\includegraphics[width=0.3\textwidth]{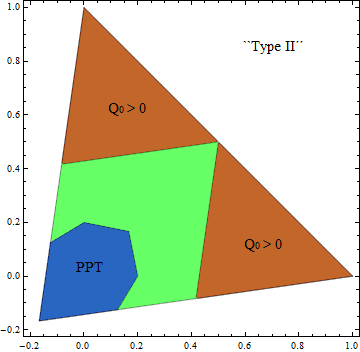}}
(c)\includegraphics[width=0.3\textwidth]{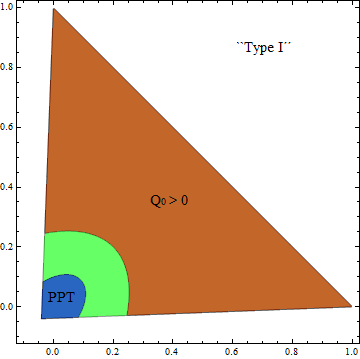}(d)\includegraphics[width=0.3\textwidth]{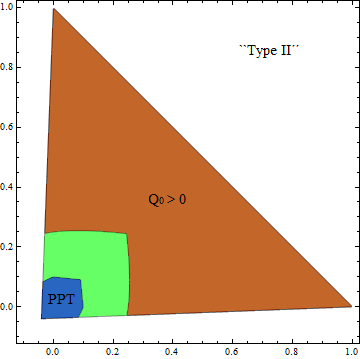}
(e)\includegraphics[width=0.3\textwidth]{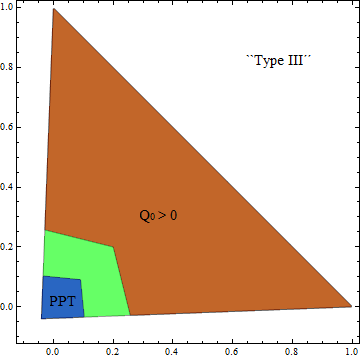}
\caption{(Color online) These pictures represents different slices through the magic simplex of entangled entanglement for three qubits (a),(b) and three qutrits (c),(d),(e), i.e. each colored point corresponds to a density matrix. On the axes the parameters $\alpha,\beta$ are plotted, respectively. Positivity constrains the parameter in a triangular area. The blue (PPT) area corresponds to positive eigenvalues under partial transposition which depends on the chosen mixture of the GHZ states, for qubits Eq.~(\ref{typesqubits}) and for qutrits Eq.~(\ref{typesqutrits}). The brown ($Q_0>0$) areas correspond to states which are detected by criterion $Q_{GHZ}$ sensitive to GHZ-type of genuine multi-partite entanglement. While the geometry of positivity is independent on the chosen mixture of GHZ states, entanglement in general and GHZ-type of genuine multi-partite entanglement are in general not. In particular, more degrees of freedoms allow for different geometries reflecting new potential physical properties.}
\label{fig_trianular1}\end{figure}

\subsubsection{Uncoloured noise mixed with one GHZ states}

As a first example let us consider a three-particle GHZ state mixed with white/uncoloured noise
\beq
\sigma_{\textrm{noise}}&=& \frac{1-\alpha}{8} \mathbbm{1}^{\otimes 3}+\alpha\; \rho_{\Phi_d^3(s,k,l)}\nonumber\\
&=& \frac{1-\alpha}{8} \mathbbm{1}^{\otimes 3}+\alpha\; \mathbbm{1}\otimes W_{s,0}\otimes W_{k,l}\; \rho_{\Phi_d^3(0,0,0)}\; \mathbbm{1}\otimes W_{s,0}^\dagger\otimes W_{k,l}^\dagger\;,
\eeq
where we used the definition
$\rho_{\Phi^d_n(s_1,s_2,\dots,k,l)}:=|\Phi^d_n(s_1,s_2,\dots,k,l)\rangle\langle \Phi^d_n(s_1,s_2,\dots,k,l)|$ and the $|\Phi^d_n(s_1,s_2,\dots,k,l)\rangle$ are given in Eq.~(\ref{generaldefinitionstates}).

Obviously, the eigenvalues of this class of states $\sigma_{\textrm{noise}}$ cannot depend on the choice of the GHZ state. The same holds true if we apply a partial transpose operation to any of the three subsystems. The positivity condition, $\sigma_{\textrm{noise}}\geq 0$, and positivity under partial transposition condition, $\sigma^{\mathcal{PT}}_{\textrm{noise}}\geq 0$, leads to the constraints
\beq
\textrm{\textit{Positivity:}}&& \frac{1}{d^3}(1 - \alpha)\geq 0 \;\&\; \frac{1}{d^3} (1 + (d^3-1) \alpha)\geq 0\\
\textrm{\textit{Positivity under partial transposition:}}&&\frac{1}{d^3}(1 - \alpha)\geq 0\;\&\;\frac{1}{d^3}(1 - (d^2+1) \alpha)\geq 0 \;\&\; \frac{1}{d^3}(1 + (d^2-1) \alpha)\geq 0.
\eeq
Thus positivity requires $\alpha\in[-\frac{1}{d^3-1},1]$ and the state $\sigma_{\textrm{noise}}$ is certainly entangled for $\alpha\in\{\frac{1}{d^2-1},1]$.

Now let us analyze the genuine multi-partite entanglement content for this class of states detected by the criterion $Q_{GHZ}(\rho)$. In the following we denote the criterion $Q_{GHZ}(\rho)$ by $Q_0$ to emphasize that the optimized separable state $|\chi\rangle$ is not necessarily in the computational basis. The only nonzero off-diagonal elements come from the contribution of GHZ state and in an optimal basis choice they correspond to $\frac{|\alpha|}{d}$. The only negative terms are due to the noise contribution, since we have for three particles three different partitions and each time two diagonal terms are multiplied (see Eq.(\ref{criteriondetailed})), thus the optimized value for $Q_0$ (for a given dimension $d$) is
\beq
Q_{0}(\sigma_{\textrm{noise}})&=&2\left(\frac{|\alpha|}{d}-3\cdot\frac{1-\alpha}{d^3}\right)\;.
\eeq
This function is positive for all $\alpha>\frac{3}{d^2+3}$ and therefore genuine multi-partite entangled. This bound differs clearly from the bound of being entangled $\alpha>\frac{1}{d^2-1}$ given by Peres-Horodecki criterion.

For any dimension $d$ and number of particles $n$ we straightforwardly obtain the result
\beq
Q_{0}(\sigma_{\textrm{noise}})&=&2\left(\frac{|\alpha|}{d}-\frac{1}{2} \left(\sum_{i=1}^{n-1} \left(\begin{array}{c} n\\i\end{array}\right)\right)\cdot\frac{1-\alpha}{d^n}\right)\;.
\eeq

\begin{figure}
	\centering
\includegraphics[width=0.7\textwidth]{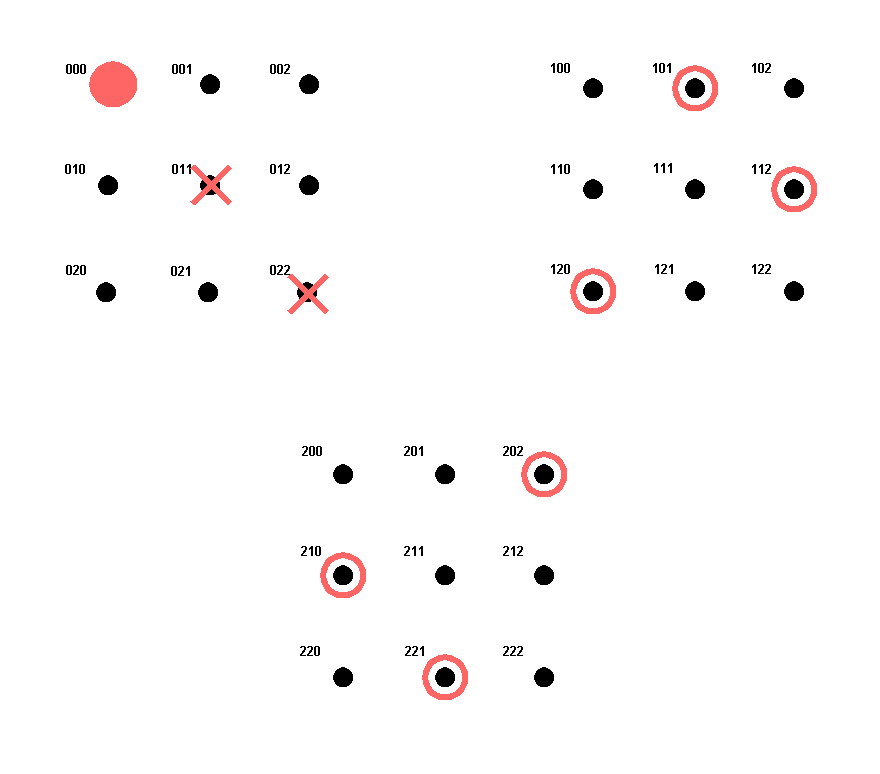}
\caption{(Color online) These drawings illustrate the phase-space structure of the magic simplex $\mathbbm{S}$ of entangled entanglement for three qutrits (compare with Fig.~\ref{fig:Qutrits}).
\label{Qutrits_type_I-III}}\end{figure}

Let us here also comment on the other criterion designed to detect $W$-type of genuine multi-partite entangled states, Eq.(\ref{Dicketripartite}). If we apply this criterion to our noisy GHZ state $\sigma_{\textrm{noise}}$ one obtains in general a lower value. For example, for tripartite qubits only states with $\alpha>0.6$ are detected to be genuine multi-partite entangled. Consequently, this criterion does detect genuine multi-partite entanglement, however, in a smaller parameter region compared to the GHZ-type of criterion. Remarkably, if one chooses the state $|\chi\rangle$ that optimize $Q_{GHZ}$ the criterion $Q_{Dicke}$ is not violated, this basis dependence can be exploited to reveal the physical difference between different types of genuine multi-partite entanglement that in consequence allows for different quantum applications.

\subsubsection{Two GHZ states mixed with noise}

Now let us consider the following class of states within the magic simplex
\beq\label{statesclass2}
\sigma(s,s';k,l;k',l')&=& \frac{1-\alpha-\beta}{8} \mathbbm{1}^{\otimes 3}+\alpha\; \rho_{\Phi_d^3(s',k',l')}+\beta\; \rho_{\Phi_d^3(s,k,l)}\nonumber\\
&=& \frac{1-\alpha-\beta}{8} \mathbbm{1}^{\otimes 3}+\alpha\; \mathbbm{1}\otimes W_{s',0}\otimes W_{k',l'}\; \rho_{\Phi_d^3(0,0,0)}\; \mathbbm{1}\otimes W_{s',0}^\dagger\otimes W_{k',l'}^\dagger\nonumber\\
&&\qquad\qquad\qquad+\beta\; \mathbbm{1}\otimes W_{s,0}\otimes W_{k,l}\;\rho_{\Phi_d^3(0,0,0)}\;\mathbbm{1}\otimes W_{s,0}^\dagger\otimes W_{k,l}^\dagger .
\eeq
Without loss of generality let us assume $s'=k'=l'=0$ and denote this class of states by $\sigma(s,k,l)$. For any non-trivial choice of the parameters, i.e. $s\not=0$ or $k\not=0$ or $l\not=0$, the positivity constraint on the density matrix leads to
\beq
\frac{1}{d^3}\left(1-\alpha-\beta\right)&\geq& 0,\nonumber\\
\frac{1}{d^3}\left(1+(d^3-1)\alpha-\beta\right)&\geq& 0,\nonumber\\
\frac{1}{d^3}\left(1-\alpha+(d^3-1)\beta\right)&\geq& 0\;.
\eeq
This corresponds to a triangle area in the $\alpha,\beta$ parameter space which can be considered as a slice through our the magic simplex and is visualized in Figs.~\ref{fig_trianular1}~(a)-(e) for tripartite qubit and qutrit states. The geometrical properties due to the positivity do not change with dimension $d$ nor as later shown with the number of particles $n$.

\begin{figure}[h!]
\flushleft{(a)\includegraphics[width=0.35\textwidth]{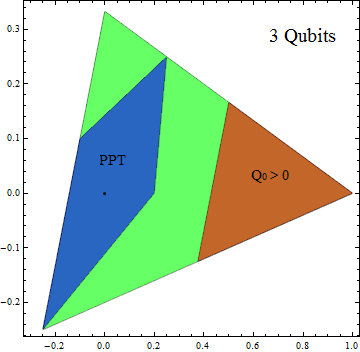}(b)\includegraphics[width=0.35\textwidth]{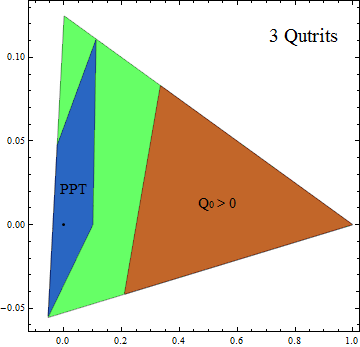}}
(c)\includegraphics[width=0.35\textwidth]{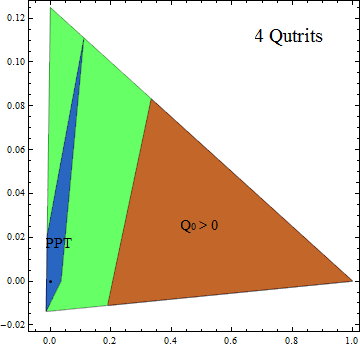}(d)\includegraphics[width=0.35\textwidth]{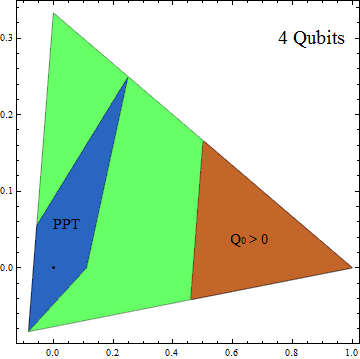}
(e)\includegraphics[width=0.35\textwidth]{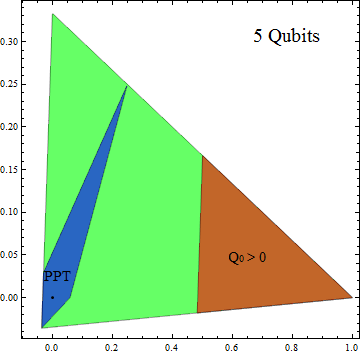}(f)\includegraphics[width=0.35\textwidth]{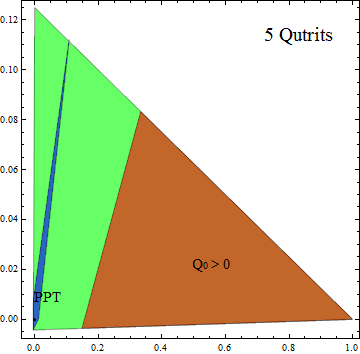}
\caption{Here we compare the geometry of entanglement properties for (a) $n=3, d=2$, (b) $n=3, d=3$, (c) $n=4, d=2$, (d) $n=4, d=3$ and (e) $n=5, d=2$ for the mixture of the unity and one GHZ states within the remaining square (i.e. $d^2-1$ states). We observe similar geometrical properties for different dimensions and number of particles. Increasing the dimension increases the parameter space of detected genuine multi-partite entanglement, whereas increasing the number of particles does not dramatically change the detected region.}
\label{fig_trianular2}\end{figure}

The Peres-Horodecki criterion, i.e. the positivity with respect to the partial transposition in one subsystem is more involved, since it depends on the choice of the parameters. We therefore start with the case of qubits $d=2$. Here we find two different types:

\beq\label{typesqubits}
&&\textrm{Type I\; (s=0;k=l=1):}\quad\sigma(s,k,l)^{\mathcal{PT}}\geq 0\nonumber\\
&&\hphantom{\textrm{Type I}\; (s=0;k=1;l=1):\quad}\Longrightarrow \min\left\{ \frac{1}{8} (1 -\alpha - 5 \beta), \frac{1}{8} (1 -
   5 \alpha - \beta), \frac{1}{8} (1 +
   3 \alpha - \beta), \frac{1}{8} (1 - \alpha + 3 \beta)\right\} \geq 0\nonumber\\
&&\textrm{Type II}\; (else):\quad\sigma(s,k,l)^{\mathcal{PT}}\geq 0\nonumber\\
&&\hphantom{\textrm{Type I}\; (s=0;k=1;l=1):\quad}\Longrightarrow \min\left\{ \frac{1}{8} (1-\alpha - \beta), \frac{1}{8} (1 + 3 \alpha - 5 \beta),\frac{1}{8} (1 - 5 \alpha + 3 \beta), \frac{1}{8} (1 + 3 \alpha + 3 \beta) \right\}\geq 0.\nonumber\\
\eeq
The first type is a mixture of GHZ states which belong to the same square and are phase flipped and shifted since then contribution of both states add in a single off-diagonal element.
Both geometries are visualized in Figs.~\ref{fig_trianular1}~(a) and (b). Our next question is:\\
\\
\textit{Should genuine multi-partite entanglement depend on the choice of the mixture of GHZ states?}
\\
\\
Since partial transposition reflects entanglement we would assume it to hold also for genuine multi-partite entanglement, on the other hand genuine multi-partite entanglement reflects a property of the entanglement shared by all three parties which could wipe out this effect. Considering the criterion $Q_{0}$ we expect two different results: one for the case when only one off-diagonal element of the density matrix has a contribution from one GHZ state or the other GHZ state (Type II), and in the other case when the off-diagonal element has contributions from both GHZ states (Type I). Consequently, we obtain (assuming without loss of generality $\alpha>\beta$):
\beq
&&\textrm{Type I}\; (s=0;k=l=1):\quad Q_{0}(\sigma)=2\left(\frac{|\alpha-\beta|}{d}-3\cdot \frac{1-\alpha-\beta}{d^3}\right)\nonumber\\
&&\textrm{Type II\; (else):}\quad Q_{0}(\sigma)=2\left(\frac{|\alpha|}{d}-2\cdot \frac{1-\alpha-\beta}{d^3}-1\cdot(\frac{1-\alpha-\beta}{d^3}+\frac{\beta}{d})\right),
\eeq
with $d=2$. Both functions are equal, i.e. the same parameter region is detected to be genuine multi-partite entangled as visualized in Figs.~\ref{fig_trianular1}~(a) and (b). Thus it reflects the fact that genuine multi-partite entanglement corresponds to a property that is shared by all subsystems.

Let us now discuss, if this property also holds in higher dimensions. For that we have to compute the Peres-Horodecki criterion for $d=3$ which results in three different types of states ($d=3$):
\beq\label{typesqutrits}
&&\textrm{Type I}\; (s=0;k=l=1,2):\quad\sigma(s,k,l)^{\mathcal{PT}}\geq 0\nonumber\\
&&\Longrightarrow \min\left\{ \frac{1}{d^3} (1-\alpha -\beta),\frac{1}{d^3} (1+(d^2-1) \alpha +(d^2-1) \beta),\frac{1}{d^3} \left(1-\alpha -\beta\pm d^2 \sqrt{\alpha ^2-\alpha  \beta +\beta ^2}\right)\right\}\geq 0\nonumber\\
&&\textrm{Type II\; (not I or III):} \quad\sigma(s,k,l)^{\mathcal{PT}}\geq 0\nonumber\\
&&\Longrightarrow \min\left\{\frac{1}{d^3} (1-\alpha -\beta),\frac{1}{d^3} (1-(d^2+1) \alpha -\beta),\frac{1}{d^3} (1-\alpha -(d^2+1) \beta),\frac{1}{d^3} (1+(d^2-1) \alpha -\beta),
\right.\nonumber\\
   &&\qquad\qquad\left.
\frac{1}{d^3}
   (1-\alpha +(d^2-1) \beta),\frac{1}{d^3} (1+(d^2-1) \alpha +(d^2-1) \beta), \frac{1}{d^3} \left(1-\alpha -\beta\pm d^2 \sqrt{\alpha^2-\alpha  \beta +\beta ^2}\right) \right\}\geq 0\nonumber\\
&&\textrm{Type III}\; (s=(l-k)\textrm{mod}\;d): \quad\sigma(s,k,l)^{\mathcal{PT}}\geq 0\nonumber\\
&&\Longrightarrow \min\left\{\frac{1}{d^3} (1-\alpha -\beta),\frac{1}{d^2} (1-(d^2+1) \alpha -\beta),\frac{1}{d^3} (1-\alpha -(d^2+1)\beta),\frac{1}{d^3} (1+(d^2-1) \alpha -\beta),\right.\nonumber\\
&&\qquad\qquad\left.\frac{1}{d^3} (1-\alpha +(d^2-1) \beta) \right\}\geq 0.
\eeq

These geometrical differences become also relevant for genuine multi-partite entanglement detected by $Q_{0}$
\beq\label{Q0qutrits}
&&\textrm{Type I}\; (s=0;k=l=1,2):\quad Q_{0}(\sigma)=2\left(\frac{|\alpha-(1-\omega)\beta|}{d}-3\cdot \frac{1-\alpha-\beta}{d^3}\right)\nonumber\\
&&\textrm{Type II\; (not I or III):} \quad Q_{0}(\sigma)=2\left(\frac{|\alpha|}{d}-2\cdot \frac{1-\alpha-\beta}{d^3}-1\cdot \sqrt{\frac{1-\alpha-\beta}{d^3} \frac{1-\alpha+ (d^2-1)\beta}{d^3}}\right)\nonumber\\
&&\textrm{Type III}\; (s=(l-k)\textrm{mod}\;d ): \quad Q_{0}(\sigma)=2\left(\frac{|\alpha|}{d}-3\cdot \frac{1-\alpha-\beta}{d^3}\right)\;,
\eeq
which shows that in higher dimensions the region of GHZ-type of genuine multi-partite entanglement depends on the chosen mixture of GHZ states since these functions are not equal. This reflects the fact that more degrees of freedom allow for more options concerning how entanglement manifests.

Mixing the totally mixed state $\frac{1}{3^3}\mathbbm{1}$ and the GHZ$000$ with any other GHZ state  (see Fig.~\ref{Qutrits_type_I-III}) leads to three different geometries concerning entanglement, Eq.~(\ref{typesqutrits}) and Eq.~(\ref{Q0qutrits}). Within the square states which are connected via a phase and spin flip (Type I)  posses a different geometry (see Fig.~\ref{fig_trianular1}~(c)). These two states are marked by (red) crosses. GHZ states marked by  big (red) circles are not within the square of GHZ$000$ and have a difference of one or two in the index between the flip and the phase operation in the second and third subsystems, respectively. They possess a different geometry (Type III, see Fig.~\ref{fig_trianular1}~(e)) than all the other possible mixtures (Type II, see Fig.~\ref{fig_trianular1}~(d)).


\subsubsection{Entanglement properties in one square}\label{sec:entanglement in dimensions}

The real beauty of the magic simplex is that it allows us also to compare the geometry in a well defined way for different dimensions and number of particles. Now we will consider the mixture of the unity with one GHZ state and all the remaining ones in the same square, i.e. $d^2-1$ other GHZ states, i.e.
\beq
\tau=\frac{1-\alpha-\beta}{d^n} \mathbbm{1}_{d}^{\otimes n}+ \sum_{k,l=0}^{d-1} \alpha_{k,l} \mathbbm{1}^{\otimes (n-1)} \otimes W_{k,l}\;\rho_{\Phi_n^d(0,0,0)}\;\mathbbm{1}^{\otimes (n-1)} \otimes W_{k,l}^\dagger
\eeq
where $\alpha_{k,l}$ equals $\alpha$ for $k=l=0$ else $\beta$. The geometry is presented in Fig.~\ref{fig_trianular2}. We observe that the region of generally entangled states increases with both the dimension $d$ and the number of particles $n$, however, the detection region of genuine multi-partite entangled states increases considerably with the dimension $d$, but not so much with the number of particles $n$.


\section{Conclusions}\label{sec:conclusions}

GHZ-type of entangled states are from their physics content fundamentally different to other (genuine) multi-partite entangled states.
They allow for example for secrete sharing protocols that would not work out for other genuine entangled states.


Furthermore, the naive concept of reality, that the three particles of the GHZ state always have well-defined local properties, fails promptly for such multi-partite entangled systems. Remarkably, depending on Alice's choice of measurement (projecting on linearly or circularly polarized photons in line 1) the reality content of the two photons in lines 2 and 3 on Bob's side
switches. According to EPR we can attribute an element of reality to the entangled Bell state on Bob's side but not to each individual photon separately. This switching phenomenon between entanglement and separability can be traced back to different factorizations of the tensor product of algebras of the three particles (or Hilbert spaces). For the transformation from separable into entangled states we derive a physical interpretation in terms of an interferometric device, see Fig.~\ref{fig:Interferometer} and Fig.~\ref{fig:GHZ-Interferometer}.

The main result is a systematic construction method to build-up an orthonormal basis of GHZ states -- the magic simplex of entangled entanglement -- by using simple unitary operations with a cyclic property, so-called Weyl operators. This construction, applicable for any number of particles and degrees of freedom, reveals an interesting cyclic geometric structure when applying flip or phase operations in a subsystem, which is depicted in Fig.~\ref{fig:GHZ states} for qubits and in Fig.~\ref{fig:Qutrits} for qutrits.

After introducing the concept of genuine multi-partite entanglement and the HMGH-framework in detail we consider particular convex combinations of GHZ states and analyze geometrically its entanglement properties. We find that the genuine multi-partite property strongly depends on the choice of mixture of GHZ states such as entanglement detected by the Peres-Horodecki criterion. This is, at the first sight, unexpected since all GHZ states differ only by one local operation and genuine multi-partite entanglement involves all degrees of freedom. It is certainly of importance for experiments since experimenters need to have knowledge about the local operations to control the final output state. These properties are the key ingredients to be exploited for any quantum algorithm or any quantum cryptography protocol. In Fig.~\ref{fig_trianular1} the geometry of qubit and qutrit case for three particles is plotted and in Fig.~\ref{fig_trianular2} the geometry of different numbers of qubits and qutrits are compared. This shows explicitly  the advantage of the magic simplexes since they allow for a defined and conceptually clear way to compare state spaces of different dimensions and number of particles.

The next step would be to exploit the revealed geometry of different mixtures of GHZ-states for quantum information theoretic applications.


\begin{acknowledgments}
BCH acknowledges gratefully the Austrian Science Fund (FWF-P23627-N16).
\end{acknowledgments}


\section*{References}


\begin{thebibliography}{10}

\bibitem{bertlmann-zeilinger02}
R. A. Bertlmann and A. Zeilinger (eds.), \emph{Quantum [Un]speakables}, Springer (2002).

\bibitem{horodecki99}
P. Horodecki, M. Horodecki, and R. Horodecki, Phys. Rev. Lett. \textbf{82}, 1056 (1999).

\bibitem{baumgartner-hiesmayr-narnhofer06}
B. Baumgartner, B. C. Hiesmayr, and H. Narnhofer, Phys. Rev. A \textbf{74}, 032327 (2006).

\bibitem{baumgartner-hiesmayr-narnhofer07}
B. Baumgartner, B. C. Hiesmayr, and H. Narnhofer, J. Phys. A: Math. Theor. \textbf{40}, 7919 (2007).

\bibitem{baumgartner-hiesmayr-narnhofer08}
B. Baumgartner, B. C. Hiesmayr, and H. Narnhofer, Phys. Lett. A \textbf{372}, 2190 (2008).

\bibitem{bertlmann-krammer-AnnPhys09}
R. A. Bertlmann and P. Krammer, Ann. Phys. \textbf{324}, 1388 (2009).

\bibitem{bertlmann-krammer-PRA-78-08}
R. A. Bertlmann and P. Krammer, Phys. Rev. A \textbf{78}, 014303 (2008).

\bibitem{bertlmann-krammer-PRA-77-08}
R. A. Bertlmann and P. Krammer, Phys. Rev. A \textbf{77}, 024303 (2008).

\bibitem{Chruscinski1}
D. Chr\'usci\'nski and G. Sarbicki, \textit{Entanglement witnesses: construction, analysis and classification}, arXiv:1402.2413.

\bibitem{Chruscinski2}
B. Bylicka, D. Chr\'usci\'nski and J. Jurkowski, J. Phys. A: Math. Theor. \textbf{46}, 205303 (2013).

\bibitem{hiesmayrloeffler1}
B. C. Hiesmayr and W. L\"offler
New J. Phys. \textbf{15}, 083036 (2013).

\bibitem{hiesmayrloeffler2}
B. C. Hiesmayr and W. L\"offler,
Phys. Scr. Vol. \textbf{2014}, 014017 (2014).


\bibitem{HHHH07}
R. Horodecki, P. Horodecki, M. Horodecki, and K. Horodecki, Rev. Mod. Phys. \textbf{81}, 865 (2009).

\bibitem{gmedet3}
O. G\"uhne and G. T\' oth, Physics Reports 474, 1 (2009).

\bibitem{TBKN}
W. Thirring, R.A. Bertlmann, P. K\" ohler, and H. Narnhofer, Eur. Phys. J. D \textbf{64}, 181 (2011).

\bibitem{zanardi2001}
P. Zanardi, Phys. Rev. Lett. \textbf{87}, 077901 (2001).

\bibitem{hmgh}
M. Huber, F. Mintert, A. Gabriel, and B.C. Hiesmayr, Phys. Rev. Lett. \textbf{104}, 210501 (2010).

\bibitem{GHZpaper1}
Ch. Eltschka and J. Siewert, Scientific Reports \textbf{2}, 942 (2012).

\bibitem{GHZpaper2}
Ch. Eltschka and J. Siewert, Phys. Rev. Lett. \textbf{108}, 020502 (2012).

\bibitem{Mario}
S. Filippov and M. Ziman, Phys. Rev. A  \textbf{88}, 032316 (2013).

\bibitem{Szalay}
S. Szalay, Phys. Rev. A  \textbf{83}, 062337 (2011).

\bibitem{Krenn-Zeilinger}
G. Krenn and A. Zeilinger, Phys. Rev. A \textbf{54}, 1793 (1996).

\bibitem{GHZ-paper}
D.M. Greenberger, M.A. Horne, and Zeilinger, \emph{Going beyong Bell's Theorem}, in \emph{Bell's Theorem, Quantum Theory, and Conceptions of the Universe}, M. Kafatos (ed.), p. 73, Kluwer Academics, Dortrecht, The Netherlands 1989.

\bibitem{GHZ-paper2}
D.M. Greenberger, M.A. Horne, and Zeilinger, Am. Journ. Phys. \textbf{58}, 1131 (1990).

\bibitem{Walther-Resch-Brukner-Zeilinger}
P. Walther, K.J. Resch, \v C. Brukner, and A. Zeilinger, Phys. Rev. Lett. 97, 020501 (2006).

\bibitem{neutron}
D. Erd\"osi, M. Huber, B.C. Hiesmayr and Y. Hasegawa, New J. Phys. 15, 023033 (2013).

\bibitem{bertlmann-narnhofer-thirring02}
R. A. Bertlmann, H. Narnhofer, and W. Thirring, Phys. Rev. A \textbf{66}, 032319 (2002).

\bibitem{vollbrecht-werner-PRA00}
K. G. H. Vollbrecht and R. F. Werner, Phys. Rev. A \textbf{64}, 062307 (2000).

\bibitem{horodecki-R-M96}
R. Horodecki and M. Horodecki, Phys. Rev. A \textbf{54}, 1838 (1996).

\bibitem{Uchida-DA}
G. Uchida, \emph{Geometry of GHZ-type quantum states}, Diploma thesis, University of Vienna (2013).

\bibitem{Ringelspiel-Lied}
\emph{``Sch\"on ist so ein Ringelspiel ...''}, famous Viennese song by Hermann Leopoldi about the pleasure of using the Carousel in the Viennese Prater, http://www.youtube.com/watch?v=TB1vp6QjLCs.

\bibitem{ViennesePrater}
\emph{Viennese Prater} is a very popular amusement park in Vienna.

\bibitem{Englert}
B.-G. Englert, Rev. Lett. {\bf 77}, 2154 (1996).

\bibitem{bramon}
A. Bramon, G. Garbarino, B.C. Hiesmayr,
Phys. Rev. A {\bf 69},  022112 (2004).

\bibitem{CPBohr}
M. Huber and B.C. Hiesmayr, Phys. Lett. A \textbf{372}, 3608 (2008).

\bibitem{SHH1}
Ch. Spengler, M. Huber and B. C. Hiesmayr,
J. Phys. A: Math. Theor. {\bf 43}, 385306 (2010).

\bibitem{SHH2}
Ch. Spengler, M. Huber and B. C. Hiesmayr,
J. Math. Phys. {\bf 53}, 013501 (2012).

\bibitem{ghh}
A. Gabriel, M. Huber and B.C. Hiesmayr, Quantum Information and Computation (QIC) 10, No. 9 \& 10, pp 829 (2010).

\bibitem{hmghcontinous}
A. Gabriel, M. Huber, S. Radic and B.C. Hiesmayr, Phys. Rev. A {\bf 83}, 052318 (2011).

\bibitem{Gurvits}
L. Gurvits, Journal of Computer and System Sciences {\bf 69}, 448 (2004).

\bibitem{peres}
A. Peres, Phys. Rev. Lett. {\bf 77}, 1413 (1996).

\bibitem{horodecki-M-P-R-M96}
M. Horodecki, P. Horodecki, and R. Horodecki, Physics Letters A  {\bf 223}, 1 (1996).

\bibitem{SHH}
S. Schauer, M. Huber and B.C. Hiesmayr, Phys. Rev. A {\bf 82}, 062311 (2010).

\bibitem{c1}
S.N. Filippov, A.A. Melnikov and M. Ziman
Phys. Rev. A {\bf 88}, 062328  (2013).

\bibitem{c2}
S.M.H. Rafsanjani, C.J. Broadbent and J.H. Eberly,
Phys. Rev. A {\bf 88}, 062331 (2013).

\bibitem{c3}
S.M. Giampaolo and B.C. Hiesmayr,
Phys. Rev. A {\bf 88}, 052305   (2013).

\bibitem{c4}
S.M. Giampaolo and B.C. Hiesmayr,
New J. Phys. {\bf 16}, 093033  (2014).

\bibitem{c5}
H.S. Dhar, A. Sen (De) and U. Sen,
Phys. Rev. A {\bf 111}, 070501   (2013).

\bibitem{c6}
R. Sweke, I. Sinayskiy and F. Petruccione, Journal of Physics B-Atomic Molecular and Optical Physics \textbf{46}, 104004 (2013).

\bibitem{c7}
X. Zha, C. Yuan and Y. Zhang,
Laser Physics Letters {\bf 10}, 045201 (2013).

\bibitem{c8}
A. Gabriel and B.C. Hiesmayr, European Physics Letters {\bf 101}, 30003 (2013).

\bibitem{mconcurrence}
B.C. Hiesmayr and M. Huber, Phys. Rev. A {\bf 78}, 012342 (2008).

\bibitem{DickeCriteria}
M. Huber, P. Erker, H. Schimpf, A. Gabriel and B.C. Hiesmayr, Phys. Rev. A {\bf 83}, 040301(R) (2011).

\bibitem{deviceindependGME}
A. Gabriel, L. Rudnicki and B.C. Hiesmayr
New J. Phys. {\bf 15}, 073033 (2013).

\end{thebibliography}
\end{document}